\documentclass[aps,prd,nofootinbib,showkeys,preprint,floatfix]{revtex4-1}

\usepackage{amsmath}
\usepackage{amssymb}
\usepackage{graphicx}
\usepackage{rotating}
\usepackage{color} %This is essential for placing graphics onto
\usepackage{multirow} % To use multirow command in tables
 \usepackage[T1]{fontenc}

\newcommand{\AddrAHEP}{
  {\it AHEP Group, Instituto de F\'{\i}sica Corpuscular --
    C.S.I.C./Universitat de Val{\`e}ncia \\
    Edificio de Institutos de Paterna, Apartado 22085,
  E--46071 Val{\`e}ncia, Spain}}

\newcommand{\AddrLisb}{%
 Departamento de F\'\i sica and CFTP, Instituto Superior T\'ecnico\\
Universidade T\'ecnica de Lisboa\\
          Av. Rovisco Pais 1, 1049-001 Lisboa, Portugal }

\newcommand{\AddrWur}{%
Institut f\"ur Theoretische Physik und Astronomie, 
Universit\"at W\"urzburg\\
Am Hubland, 
97074 Wuerzburg}

\def\gsim{\raise0.3ex\hbox{$\;>$\kern-0.75em\raise-1.1ex\hbox{$\sim\;$}}}
\def\lsim{\raise0.3ex\hbox{$\;<$\kern-0.75em\raise-1.1ex\hbox{$\sim\;$}}}

\begin{document}

\preprint{CFTP/11-016}  
\preprint{IFIC/11-55}  

\title{Dark matter and LHC phenomenology in a left-right 
supersymmetric model}

\author{J. N. Esteves}\email{joaomest@cftp.ist.utl.pt}
\author{J. C. Romao}\email{jorge.romao@ist.utl.pt}\affiliation{\AddrLisb}
\author{M. Hirsch} \email{mahirsch@ific.uv.es}\affiliation{\AddrAHEP}
\author{W. Porod} \email{porod@physik.uni-wuerzburg.de}
\author{F. Staub}\email{florian.staub@physik.uni-wuerzburg.de}
\author{A. Vicente}\email{avelino.vicente@physik.uni-wuerzburg.de}
\affiliation{\AddrWur}



\begin{abstract}
Left-right symmetric extensions of the Minimal Supersymmetric Standard
Model can explain neutrino data and have potentially interesting
phenomenology beyond that found in minimal SUSY seesaw models. Here we
study a SUSY model in which the left-right symmetry is broken by
triplets at a high scale, but significantly below the GUT
scale. Sparticle spectra in this model differ from the usual
constrained MSSM expectations and these changes affect the relic
abundance of the lightest neutralino. We discuss changes for the
standard stau (and stop) co-annihilation, the Higgs funnel and the
focus point regions. The model has potentially large lepton flavour
violation in both, left and right, scalar leptons and thus allows, in
principle, also for flavoured co-annihilation. We also discuss lepton
flavour signals due to violating decays of the second lightest
neutralino at the LHC, which can be as large as $20$ fb$^{-1}$ at
$\sqrt{s}=14$~TeV.
\end{abstract}

\maketitle

%\tableofcontents

\section{Introduction}

Left-right (LR) symmetric extensions of the MSSM (``Minimal 
Supersymmetric extension of the Standard Model'') automatically 
contain the correct ingredients to explain the observed neutrino 
masses and mixings. The right-handed neutrino superfield, $\nu^c$, 
is necessarily part of the theory and breaking the LR symmetry 
by $SU(2)_R$ triplets generates at the same time a Majorana mass 
term for the $\nu^c$ and thus a seesaw mechanism 
\cite{Minkowski:1977sc,MohSen,Schechter:1980gr,Cheng:1980qt}, 
in contrast to type-I seesaw models where a Majorana mass term is 
merely added by hand. In LR models new gauge 
and Higgs fields appear with masses below the GUT scale, which 
change the running of all parameters under 
RGE evolution. In particular, the change in the running of the 
soft supersymmetry breaking masses can lead to potentially 
interesting effects in the phenomenology of SUSY models, even if 
the scale of LR breaking is above the energy range testable by 
accelerator experiments. 

Quite a large number of different LR models have been discussed 
in the literature. The original (non-supersymmetric) LR models 
\cite{Pati:1974yy,Mohapatra:1974gc,Senjanovic:1975rk} break the 
$SU(3)_C\times SU(2)_L \times SU(2)_R \times U(1)_{B-L}$ group 
by scalar doublets. Later it was realized that breaking the LR 
symmetry by (a pair of) triplets with $B-L=2$ generates automatically 
also a seesaw mechanism \cite{MohSen,Mohapatra:1980yp}. Four 
triplets are needed \cite{Cvetic:1983su} in the supersymmetric 
version of this LR model: $\Delta(1,3,1,2)$ and $\Delta^c(1,1,3,-2)$ 
due to the LR symmetry and $\bar{\Delta}(1,3,1,-2)$ and 
$\bar{\Delta}^c(1,1,3,2)$ for anomaly cancellation. Aulakh et al. 
\cite{Aulakh:1997ba,Aulakh:1997fq} extended this minimal LR model 
\footnote{Several different realizations of LR models have been 
called minimal in the literature. These contain such diverse 
variants as the models of \cite{FileviezPerez:2008sx} and 
\cite{Brahmachari:2003wv,Siringo:2003hh}. While in the SUSY model of 
\cite{FileviezPerez:2008sx} only one bi-doublet is introduced and $B-L$ 
(and thus also R-parity) is broken by the vacuum expectations value 
of the ${\tilde \nu^c}$, the non-SUSY models of 
\cite{Brahmachari:2003wv,Siringo:2003hh} do not introduce any bi-doublet, 
but only a pair of left- and right- doublets. In this construction fermion 
masses have to arise from non-renormalizable operators 
\cite{Brahmachari:2003wv,Siringo:2003hh}. }
introducing an additional pair of triplets $\Omega(1,3,1,0)$ and 
$\Omega^c(1,1,3,0)$ with zero lepton number. The advantages of this 
setup, in the following called the $\Omega$LR model, are two-fold 
\cite{Aulakh:1997fq}: First, the LR breaking 
minimum leaves R-parity unbroken already at tree level and, second, 
a non-trivial CKM matrix for quarks is generated easily and without 
resorting to flavour violating soft terms \cite{Babu:1998tm}. 

Both, the minimal SUSY LR model \cite{Cvetic:1983su} as well as the 
$\Omega$LR model \cite{Aulakh:1997ba,Aulakh:1997fq} break the LR 
symmetry at an energy scale far above the range accessible to 
accelerator experiments. Only indirect tests of these models are 
therefore possible, all of which require some assumptions about 
the high scale boundary conditions for the soft SUSY breaking terms. 
Assuming CMSSM (``constrained'' MSSM) \cite{Drees:2004jm} boundary 
conditions, two kind of indirect signals are possible, in principle: 
(i) Lepton flavor violating (LFV) decays and (ii) changes in the SUSY 
particle mass spectra. 

LFV decays are induced in supersymmetric models, even for strictly 
flavour blind boundary conditions, in the RGE running of the soft 
parameters, as has been shown for the case of type-I seesaw already 
in \cite{Borzumati:1986qx}. With LFV observed in neutrino oscillation 
experiments \cite{Fukuda:1998mi}, one expects LFV to occur also in 
the charged lepton sector in practically all SUSY models. However, 
there is a qualitative difference between seesaw models and LR models: 
whereas in SUSY seesaw  LFV is expected to occur dominantly in the left 
slepton sector \cite{Hisano:1995nq,Hisano:1995cp}, an LR 
symmetry implies that left and right sleptons should have equal 
soft mass terms. Even with LR symmetry then broken at low energies, 
LFV in the right slepton sector can be sizable, as was shown 
in \cite{Esteves:2010si}. Right slepton LFV could, in principle, be 
detected directly at accelerators or indirectly by measuring 
polarization in the decay $\mu \to e  \gamma$ \cite{Kuno:1999jp}. 

Potentially measurable differences in SUSY mass spectra with respect 
to CMSSM expectations can occur in seesaw type-II 
\cite{Hirsch:2008gh,Hirsch:2011cw} and type-III 
\cite{Hirsch:2011cw,Esteves:2010ff}, as well as in the $\Omega$LR 
model \cite{Esteves:2010si}. Changes with respect to CMSSM can 
best be understood analytically by forming certain ``invariants'', 
i.e. soft SUSY breaking mass parameter combinations which are to 
1-loop leading-log order constants over large ranges in CMSSM  
space \cite{Buckley:2006nv}. Although there are quantitatively 
important 2-loop corrections to these invariants, as we will 
show below in the $\Omega$LR model the invariants have {\em qualitatively} 
different behavior from all seesaw models, due to the LR symmetry. 
This is similar to the situation discussed recently in 
\cite{DeRomeri:2011ie} in the context of different $SO(10)$ based 
models. 

Even slight changes in the SUSY spectra can lead to rather sizable 
changes in the calculated relic density of the lightest neutralino, 
$\Omega_{\tilde \chi^0_1}h^2$, as has been shown for seesaw type-II 
\cite{Esteves:2009qr} and type-III \cite{Esteves:2010ff,Biggio:2010me} 
and also for an SO(10) based model \cite{Drees:2008tc}. 
This fact can be easily understood taking into account that all 
solutions in CMSSM parameter space \cite{Drees:1992am}, which 
survive the latest WMAP constraints \cite{Komatsu:2010fb}, require 
some special relations among SUSY masses in order to get a low 
enough $\Omega_{\tilde \chi^0_1}h^2$. For example, in the stau co-annihilation 
region $\Omega_{\tilde \chi^0_1}h^2$ is essentially determined by 
$\Delta m= m_{\tilde\tau_1}-m_{\tilde \chi^0_1}$, as long as $\Delta m 
\lsim 10$ GeV or so. Changing $m_{\tilde\tau_1}$ and/or $m_{\tilde \chi^0_1}$ 
by only a few GeV will then change $\Omega_{\tilde \chi^0_1}h^2$ by a large 
factor. Below we will discuss how the standard solutions to the 
DM problem are changed within the $\Omega$LR, with respect to the 
``pure'' CMSSM allowed regions. Since the $\Omega$LR model has potentially 
large LFV in the right slepton sector, the DM constraint can also be 
fulfilled using the flavoured co-annihilation solution, recently 
discussed in \cite{Choudhury:2011um}. We provide and discuss a few 
examples where flavoured co-annihilation can be realized in the 
parameter space of the $\Omega$LR model.

Finally, LFV SUSY decays might be seen at the LHC. In \cite{Esteves:2010si} 
we have shown that branching ratios 
$\tilde \chi^0_2 \to l_i l_j \tilde \chi^0_1$, 
with $i \ne j$ are potentially large within the $\Omega$LR. Here we 
extend this work by calculating the SUSY production cross section 
for $\tilde \chi^0_2$ from cascade decays, to estimate the number 
of events that could be found at the LHC. We have found  
$\sigma\times Br(\tilde \chi^0_2 \to l_i l_j \tilde \chi^0_1)$ 
up to $(10-20)$ $fb$.

The rest of this paper is organized as follows. In the next 
section, we define the basics of the $\Omega$LR model. In section 
\ref{sec:gsl} we discuss gauge coupling unification, generalities 
for the expected changes in the SUSY spectra with respect to CMSSM 
expectations and certain aspects of lepton flavour violation. 
In section \ref{sec:dm} we discuss dark matter in the $\Omega$LR model, 
while section \ref{sec:lhc} discusses lepton flavour violation 
phenomenology for the LHC. We then close with a short summary.

\section{Model} \label{sec:model}

In this section we present the model originally defined in
\cite{Aulakh:1997ba,Aulakh:1997fq}, where a two step breaking of the LR
symmetry was proposed in order to cure the potential problems related to 
the conservation of R-parity at low energies
\cite{Kuchimanchi:1993jg,Kuchimanchi:1995vk}. For further details 
see \cite{Esteves:2010si}.  

\subsubsection{Step 1: From GUT scale to $SU(2)_R$ breaking scale}

Below the GUT scale\footnote{See subsection \ref{subsec:gauge} for
more details about gauge coupling unification and the GUT scale.} the
gauge group of the model is $SU(3)_c \times SU(2)_L \times SU(2)_R
\times U(1)_{B-L}$. In addition, parity is assumed to be
conserved. Besides the quark and lepton superfields of the MSSM with
the addition of (three) right-handed neutrino(s) $\nu^c$, some
additional superfields are required to break the LR symmetry down to
the standard model gauge group. First, two generations of $\Phi$
superfields, bidoublets under $SU(2)_L \times SU(2)_R$, are
introduced. They contain the standard $H_d$ and $H_u$ MSSM Higgs
doublets. Note, that two copies are needed in order to generate a
non-trivial CKM matrix at tree-level. Furthermore, triplets under (one
of) the $SU(2)$ gauge groups are added whose gauge quantum number are given
in Table~\ref{tab:particles-step1}. Note that the model contains 
$B-L = \pm 2$ and $B-L = 0$ triplets.

With these representations, the most general superpotential compatible 
with the symmetries is
\begin{eqnarray} \label{eq:Wsuppot1}
{\cal W} &=& Y_Q Q \Phi Q^c 
          +  Y_L L \Phi L^c 
          - \frac{\mu}{2} \Phi \Phi
          +  f L \Delta L
          +  f^* L^c \Delta^c L^c \nonumber \\
         &+& a \Delta \Omega \bar{\Delta}
          +  a^* \Delta^c \Omega^c \bar{\Delta}^c
          + \alpha \Omega \Phi \Phi
          +  \alpha^* \Omega^c \Phi \Phi \nonumber \\
         &+& M_\Delta \Delta \bar{\Delta}
          +  M_\Delta^* \Delta^c \bar{\Delta}^c
          +  M_\Omega \Omega \Omega
          +  M_\Omega^* \Omega^c \Omega^c \thickspace.
\end{eqnarray}

Note that the superpotential in eq.~\eqref{eq:Wsuppot1} is invariant under
the parity transformations $Q \leftrightarrow (Q^c)^*$, $L
\leftrightarrow (L^c)^*$, $\Phi \leftrightarrow \Phi^\dagger$, $\Delta
\leftrightarrow (\Delta^c)^*$, $\bar{\Delta} \leftrightarrow
(\bar{\Delta}^c)^*$, $\Omega \leftrightarrow (\Omega^c)^*$. This
discrete symmetry fixes, for example, the $L^c \Delta^c L^c$ coupling
to be $f^*$, the complex conjugate of the $L \Delta L$ coupling, thus
reducing the number of free parameters of the model.

\begin{table}
\centering
\begin{tabular}{c c c c c}
\hline
Superfield & $SU(3)_c$ & $SU(2)_L$ & $SU(2)_R$ & $U(1)_{B-L}$ \\
\hline
$\Delta$ & 1 & 3 & 1 & 2 \\
$\bar{\Delta}$ & 1 & 3 & 1 & -2 \\
$\Delta^c$ & 1 & 1 & 3 & -2 \\
$\bar{\Delta}^c$ & 1 & 1 & 3 & 2 \\
$\Omega$ & 1 & 3 & 1 & 0 \\
$\Omega^c$ & 1 & 1 & 3 & 0 \\
\hline
\end{tabular}
\caption{Summary of the triplets of the $\Omega$LR model above the 
$SU(2)_R$ breaking scale.}
\label{tab:particles-step1}
\end{table}

Family and gauge indices have been omitted in eq.~\eqref{eq:Wsuppot1},
more detailed expressions can be found in
\cite{Aulakh:1997ba}. Moreover, the soft terms of the model can be
found in \cite{Esteves:2010si}. The LR symmetry itself does not, of
course, fix the values of the soft SUSY breaking terms. In our
numerical evaluation we will resort to CMSSM-like boundary conditions
at the GUT scale and obtain their values at the SUSY scale by means of
the RGEs of the model. Finally, the superpotential couplings 
$Y_Q$ and $Y_L$ are fixed by the low-scale standard model fermion
masses and mixing angles.

The breaking of the LR gauge group to the MSSM gauge group takes place
in two steps: $SU(2)_R \times U(1)_{B-L} \rightarrow U(1)_R \times
U(1)_{B-L} \rightarrow U(1)_Y$. In the first step the neutral
component of the triplet $\Omega^c$ takes a VEV (vacuum expectation
value):
\begin{equation}
\langle \Omega^{c \: 0} \rangle = \frac{v_R}{\sqrt{2}}
\end{equation}
which breaks $SU(2)_R$. However, since $I_{3R} (\Omega^{c \: 0}) = 0$ 
there is a $U(1)_R$ symmetry left over. Next, the group 
$U(1)_R \times U(1)_{B-L}$ is broken by
\begin{equation}
\langle \Delta^{c \: 0} \rangle = \frac{v_{BL}}{\sqrt{2}} \thickspace, \qquad 
\langle \bar{\Delta}^{c \: 0} \rangle = \frac{\bar{v}_{BL}}{\sqrt{2}} \thickspace.
\end{equation}
The remaining symmetry is now $U(1)_Y$ with hypercharge defined as
$Y = I_{3R} + \frac{B-L}{2}$. 

The supersymmetric nature of the model is very relevant for the
structure of the tadpole equations. Contrary to LR models without
supersymmetry, the tadpole equations do not link $\Omega^c$,
$\Delta^c$ and $\bar{\Delta}^c$ with their left-handed
counterparts. Thus, the left-handed triplets can have vanishing VEVs
\cite{Aulakh:1997ba} and the model leads only to a type-I seesaw.

Neglecting the soft SUSY breaking terms and the electroweak symmetry breaking
VEVs $v_d$ and $v_u$ (numerically irrelevant at this stage), and taking $v_{BL}
= \bar{v}_{BL}$, one finds the following solutions for the tadpole equations
\begin{equation} \label{tadpolesol}
v_R = \frac{2 M_\Delta}{a} \thickspace, \qquad v_{BL} 
= \frac{2}{a} (2 M_\Delta M_\Omega)^{1/2} \thickspace.
\end{equation}
This implies that the hierarchy $v_{BL}
\ll v_R$ requires $M_\Delta \gg M_\Omega$, as already discussed in
\cite{Aulakh:1997ba}.

\subsubsection{Step 2: From $SU(2)_R$ breaking scale to $U(1)_{B-L}$ 
breaking scale}

After the breaking of the LR gauge group to $SU(3)_c \times SU(2)_L
\times U(1)_R \times U(1)_{B-L}$ some particles get large masses and
decouple from the spectrum. In the case of the $\Delta$ triplets only
the neutral components of the $SU(2)_R$ triplets, $\Delta^{c \:
0}$ and $\bar{\Delta}^{c \: 0}$, remain light. The charged components
of the $\Omega^c$ triplet get masses of the order of $v_R$ and thus
only the  $SU(2)_L$ triplet $\Omega$ and the neutral superfield
$\Omega^{c \: 0}$ stay in the particle spectrum 
\cite{Aulakh:1997fq,Esteves:2010si}.

Similarly, the two bidoublet generations, $\Phi_1$ and $\Phi_2$, get
split into four $SU(2)_L$ doublets. Two of them remain light, being
identified with the two Higgs doublets of the MSSM, whereas the other
two get masses of order $v_R$. As a result, the light
Higgs doublets are admixtures of the corresponding flavour eigenstates,
their couplings to quarks and leptons being combinations of the original
Yukawa couplings, which has a strong impact on the low energy 
phenomenology.

The superpotential terms mixing the four $SU(2)_L$ doublets can be
written as ${\cal W}_M = (H_d^f)^T M_H H_u^f$, where $H_d^f = ( H_d^1,
H_d^2)$ and $H_u^f = ( H_u^1, H_u^2)$ are the \emph{flavour
eigenstates}. In order to compute the resulting couplings for the
light Higgs doublets one must rotate the original fields into their
mass basis. Since $M_H$ is not a symmetric matrix (unless $\alpha_{12}
=0$, see \cite{Esteves:2010si}) one has to rotate independently
$H_d^f$ and $H_u^f$, i.e. $H_d^f = D H_d^m$, $H_u^f = U H_u^m$. Here
$D$ and $U$ are unitary matrices and $H_d^m = ( H_d^l, H_d^h)$ and
$H_u^m = ( H_u^l, H_u^h)$ are the \emph{mass eigenstates}, with masses
$m(H_{d,u}^l) \sim 0$ (neglecting soft SUSY breaking terms and
electroweak VEVs) and $m(H_{d,u}^h) \sim v_R$.

The $D$ and $U$ rotation matrices are, in general, different. We can
parametrize them as
\begin{equation}
D = \left( \begin{array}{c c}
\cos \theta_1 & \sin \theta_1 \\
- \sin \theta_1 & \cos \theta_1
\end{array} \right) \thickspace, \qquad U = \left( \begin{array}{c c}
\cos \theta_2 & \sin \theta_2 \\
- \sin \theta_2 & \cos \theta_2
\end{array} \right)
\end{equation}
The angles $\theta_1$ and $\theta_2$ have a very strong impact on the
low energy phenomenology. This can be easily understood from the
matching conditions at the $SU(2)_R$ breaking scale. These include
\begin{eqnarray} \label{matching-1}
Y_Q^1 = \frac{1}{s_{21}} (\sin \theta_2 Y_d + \sin \theta_1 Y_u) \thickspace,
&\qquad& Y_Q^2 = \frac{1}{s_{21}} (\cos \theta_2 Y_d + \cos \theta_1 Y_u) \thickspace, \\ \label{matching-2}
Y_L^1 = \frac{1}{s_{21}} (\sin \theta_2 Y_e + \sin \theta_1 Y_\nu) \thickspace,
&\qquad& Y_L^2 = \frac{1}{s_{21}} (\cos \theta_2 Y_e + \cos \theta_1 Y_\nu) \thickspace.
\end{eqnarray}
where $s_{21} = \sin (\theta_2 - \theta_1)$. Eqs.~\eqref{matching-1}
and \eqref{matching-2} show that in the limit $\theta_1 \to \theta_2$
the high energy Yukawas $Y_Q^{1,2}$ and $Y_L^{1,2}$ diverge. This
implies that when $\theta_1$ and $\theta_2$ take similar values,
lepton flavour violating effects at low energies, induced by RGE
evolution by these Yukawa couplings, get larger.  See section
\ref{sec:low-energy} for numerical results on this issue.

Finally, neutrino masses are generated after the breaking of
$U(1)_{B-L}$ through a type-I seesaw mechanism. Note that the
superpotential in eq.~\eqref{eq:Wsuppot1} contains the term $f^* L^c
\Delta^c L^c$ which, after $U(1)_{B-L}$ gets broken, leads to $f_c^1
\nu^c \nu^c \Delta^{c \: 0}$, with $f_c^1 = - f^*$. Therefore, the
matrix $f_c^1$ leads to Majorana masses for the right-handed neutrinos
once $\Delta^{c \: 0}$ gets a VEV. We define the
seesaw scale as the lightest eigenvalue of the matrix $M_S \equiv
f_c^1 v_{BL}$.

\section{Gauge sector, spectrum and lepton flavour violation} \label{sec:gsl}

In this section we discuss some important properties of the gauge sector of the model, the 
supersymmetric spectrum and the role of the bidoublet mixing angles 
for the size of the lepton
flavour violating signatures.

\subsection{Gauge sector} \label{subsec:gauge}

Despite the conceptual advantages of LR models, LR has fallen somewhat out 
of favor within the supersymmetric community. This is most likely 
due to the fact that within the MSSM gauge coupling unification is 
achieved automatically, if the scale of SUSY particles is of the 
order of ${\cal O}(1)$ TeV or less. Additional intermediate scales tend to
destroy this nice feature unless new particles and/or threshold effects are
considered, see for example \cite{Kopp:2009xt}.

\begin{figure}
\begin{center}
\vspace{5mm}
\includegraphics[width=0.49\textwidth]{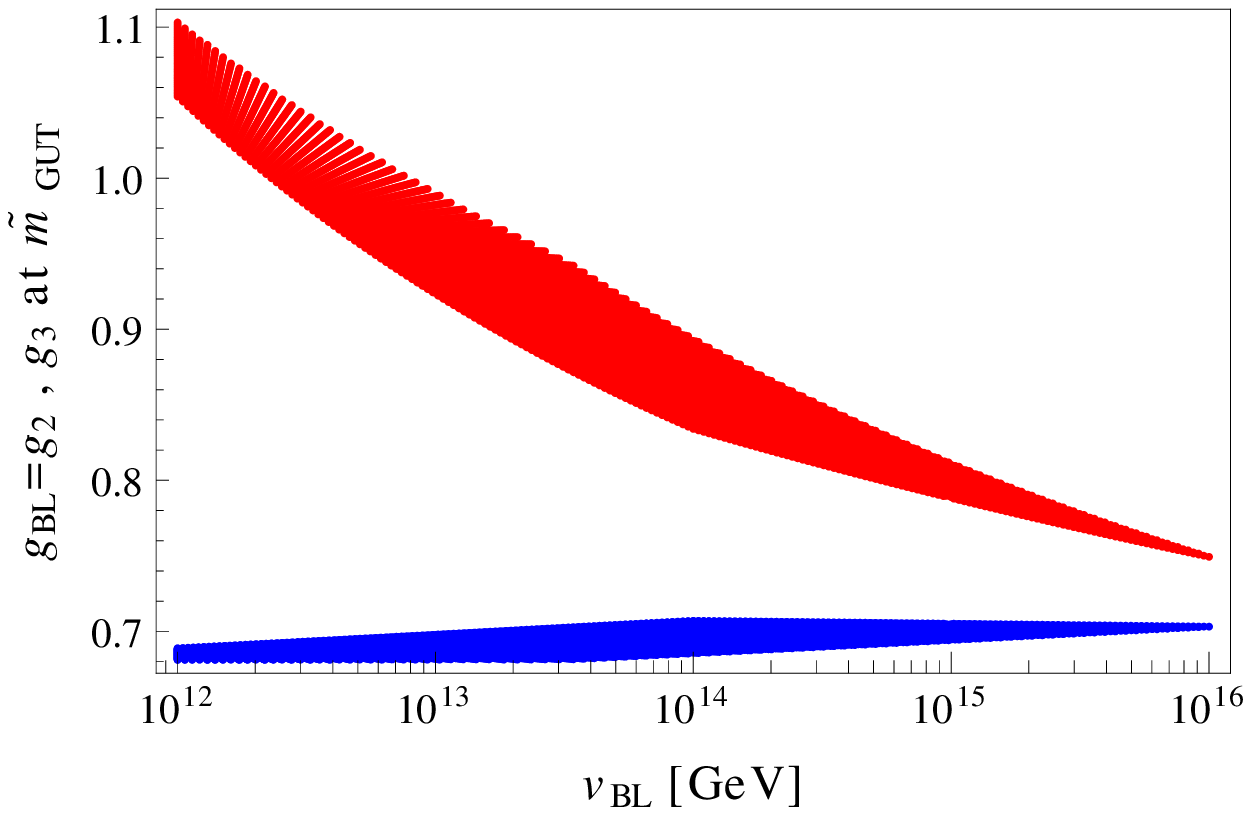}
\includegraphics[width=0.49\textwidth]{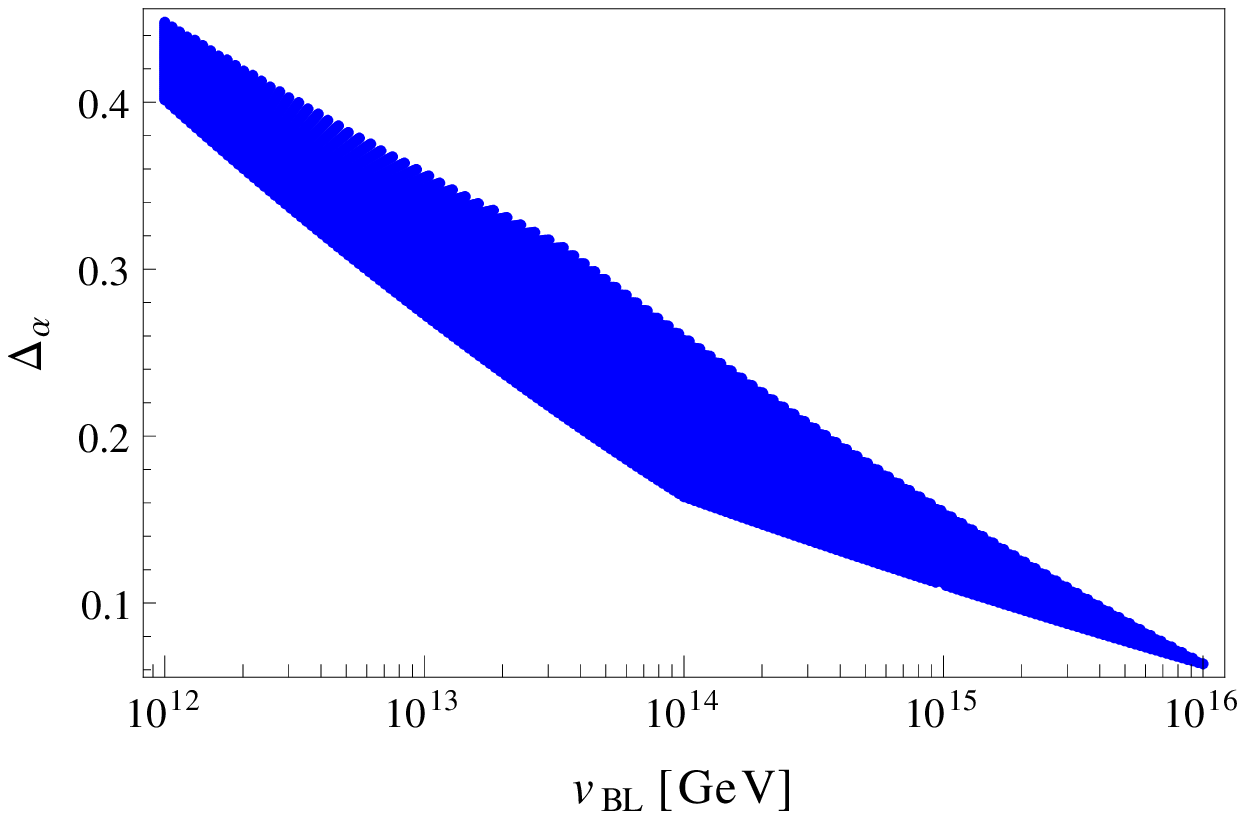}
\end{center}
\vspace{-5mm}
\caption{To the left, the gauge couplings $g_{BL} = g_2$ (blue) and $g_3$ (red)
at $\tilde{m}_{GUT}$ as a function of $v_{BL}$. To the right, $\Delta_\alpha$ as a
function of $v_{BL}$. In both figures $v_R$ is taken in the range
$[10^{14},10^{16}]$ GeV and $\tilde{m}_{GUT}$ is defined as the scale at which
$g_{BL} = g_2$.}
\label{fig:gcu-vBL}
\end{figure}

This issue is also present in the $\Omega$LR model. Let us define the scale
$\tilde{m}_{GUT}$ as the scale at which $g_{BL} = g_2$. In general, the strong
gauge coupling $g_3$ does not unify with $g_{BL} = g_2$ at $\tilde{m}_{GUT}$ and
gauge coupling unification is not fully obtained. At 1-loop one finds:
\begin{eqnarray}
\alpha_{BL}^{-1}(\tilde{m}_{GUT}) - \alpha_3^{-1}(\tilde{m}_{GUT}) &=& \frac{87}{32} \alpha_2^{-1}(m_{SUSY}) - \alpha_3^{-1}(m_{SUSY}) - \frac{55}{32} \alpha_Y^{-1}(m_{SUSY}) \nonumber \\
&& + \frac{3}{32 \pi} (29 t_{BL}+ t_R-30 t_{SUSY})
\end{eqnarray}
where $t_{SUSY} = \ln \left( m_{SUSY}/m_Z \right)$, $t_{BL} = \ln \left( v_{BL}/m_Z \right)$ and $t_R = \ln \left( v_R/m_Z \right)$. Therefore, $v_{BL}$ is
the most relevant scale for the determination of the difference between the
gauge couplings at $\tilde{m}_{GUT}$. 
This can be seen on the left side of Figure~\ref{fig:gcu-vBL},
where the values of the three gauge couplings are shown as a function of
$v_{BL}$ for $v_R \in [10^{14},10^{16}]$ GeV. For low values of $v_{BL}$ the
running of $g_{BL}$ and $g_2$ is too strong and their values at high energies
clearly depart from the one obtained for $g_3$. Furthermore, it is useful to
quantify the deviation from unification by defining
\begin{equation}
\Delta_\alpha = \frac{\alpha_{BL}(\tilde{m}_{GUT}) - \alpha_3(\tilde{m}_{GUT})}{\alpha_{BL}(\tilde{m}_{GUT}) + \alpha_3(\tilde{m}_{GUT})}.
\end{equation}
The right side of Figure~\ref{fig:gcu-vBL} shows that for $v_{BL} = 10^{12}$
GeV the parameter $\Delta_\alpha$ can be as large as 0.45. Note,
however, that this is not a principal problem.

\begin{figure}
\begin{center}
\vspace{5mm}
\includegraphics[width=0.5\textwidth]{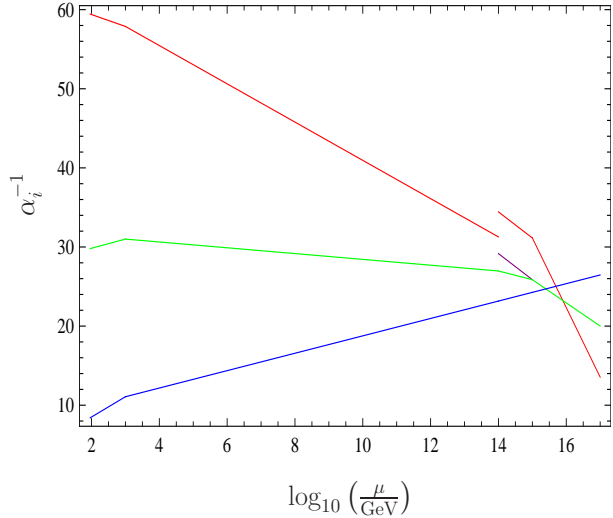}
\end{center}
\vspace{-5mm}
\caption{1-loop running of the gauge couplings for the choice of scales $m_{SUSY} = 1$ TeV, $v_{BL} = 10^{14}$ GeV and $v_R = 10^{15}$ GeV. The dependence of $\alpha_i^{-1}$, where $\alpha_i = \frac{g_i^2}{4 \pi}$, on the energy scale $\mu$ is shown. Different gauge couplings are represented in the different energy regimes. For $\mu \in \left[m_Z,v_{BL}\right]$ one has $\alpha_3^{-1}$ (blue), $\alpha_L^{-1}$ (green) and $\alpha_Y^{-1}$ (red). For $\mu \in \left[v_{BL},v_R\right]$ one has $\alpha_3^{-1}$ (blue), $\alpha_L^{-1}$ (green), $\alpha_R^{-1}$ (purple) and $\alpha_{BL}^{-1}$ (red). For $\mu > v_R$ one has $\alpha_3^{-1}$ (blue), $\alpha_2^{-1} \equiv \alpha_L^{-1} = \alpha_R^{-1}$ (green) and $\alpha_{BL}^{-1}$ (red).}
\label{fig:gcu-original}
\end{figure}

\begin{figure}
\begin{center}
\vspace{5mm}
\includegraphics[width=0.49\textwidth]{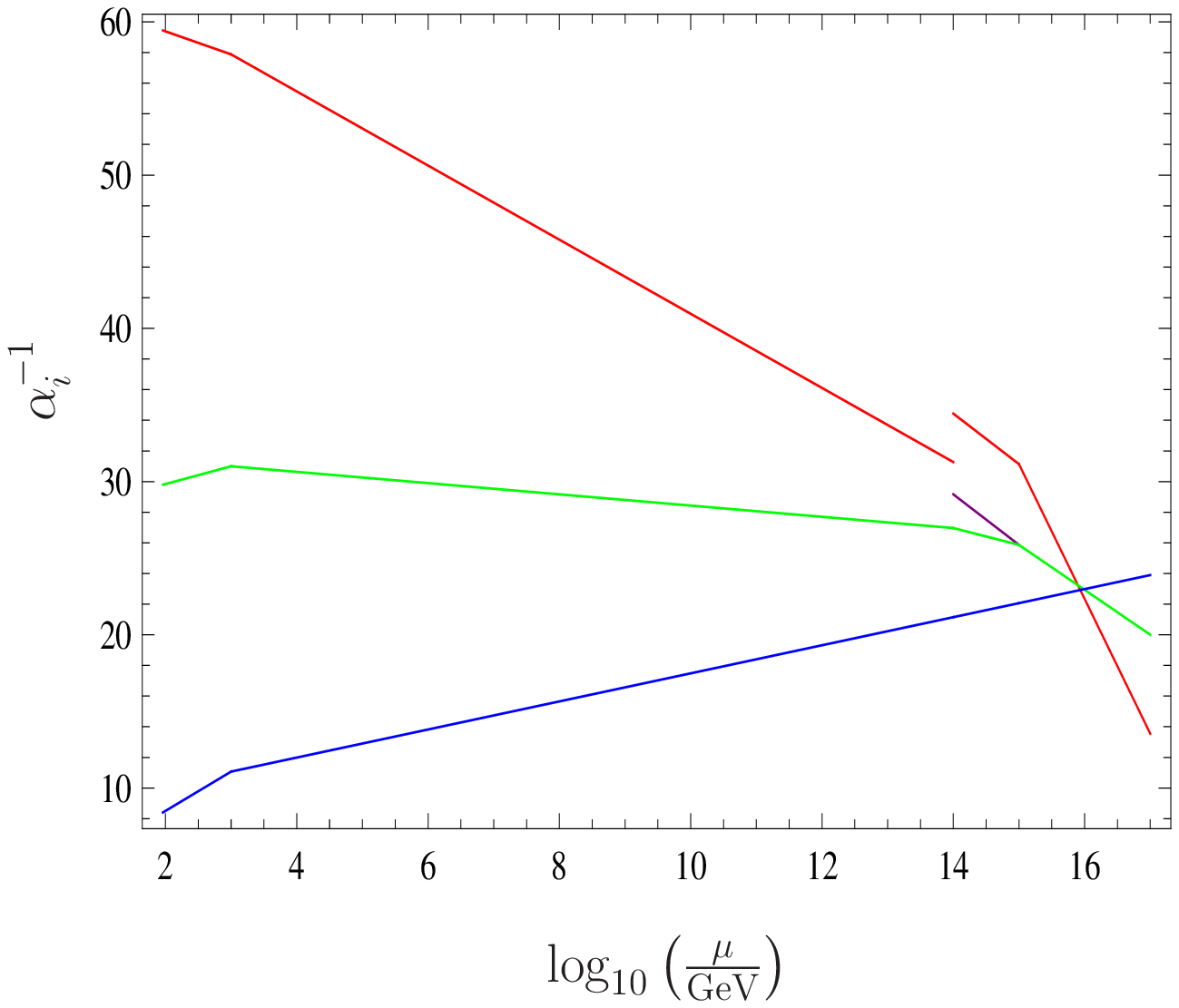}
\includegraphics[width=0.49\textwidth]{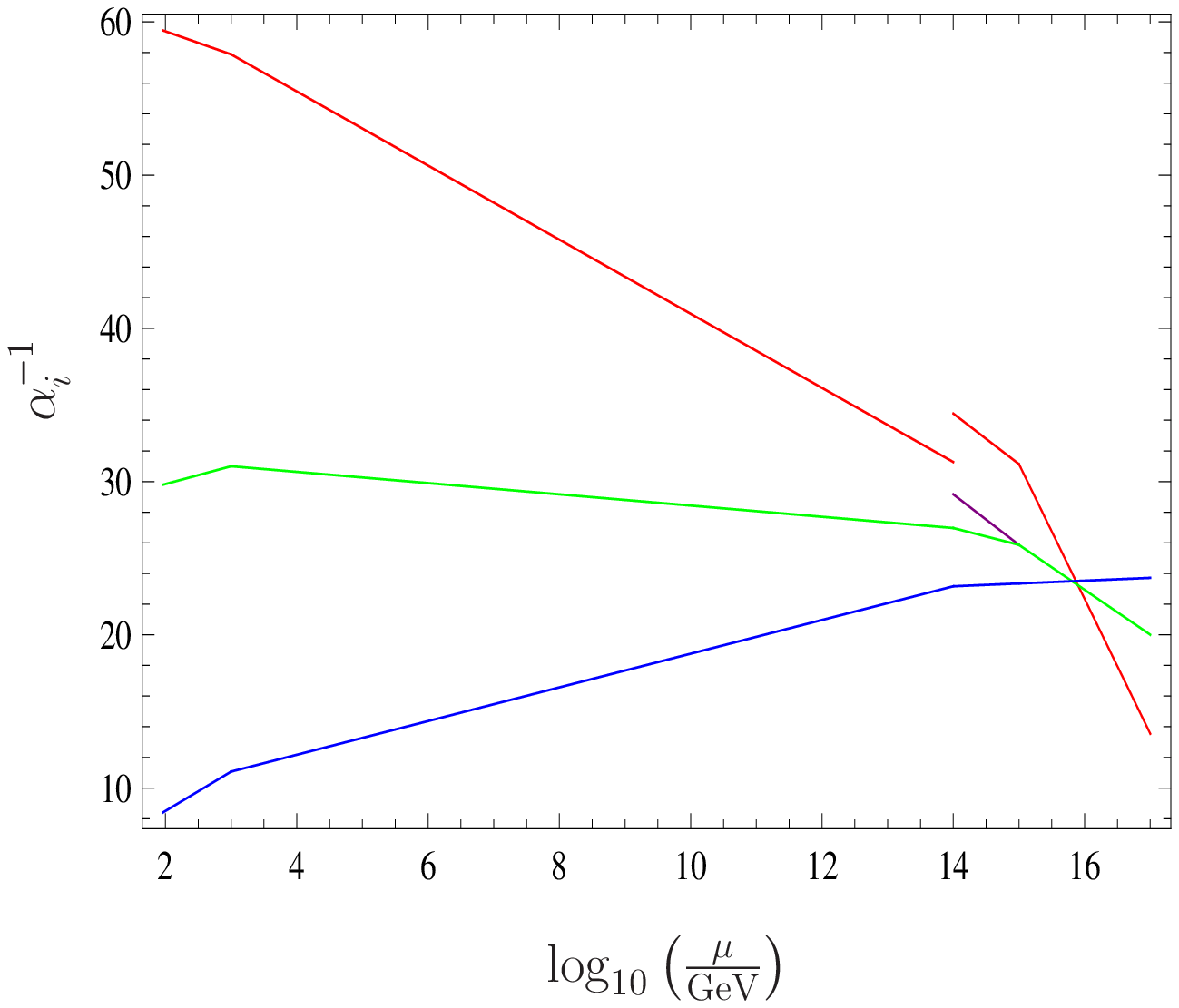}
\end{center}
\vspace{-5mm}
\caption{1-loop running of the gauge couplings for the choice of scales
$m_{SUSY} = 1$ TeV, $v_{BL} = 10^{14}$ GeV and $v_R = 10^{15}$ GeV. Contrary to
Figure~\ref{fig:gcu-original}, gauge coupling unification is obtained thanks to
additional colored superfields. In the left panel, one triplet under $SU(3)_c$,
singlet under the other gauge subgroups, is added at $m_{SUSY}$, whereas in the
right panel five generations of the same superfield are added at $v_{BL}$. See
Figure~\ref{fig:gcu-original} for the color code.}
\label{fig:gcu-solved}
\end{figure}

For instance, one possibility to recover gauge unification
 is the addition of new particles to the spectrum. As
clearly seen in Figure~\ref{fig:gcu-original}, unification is not
obtained due to a too fast running of $g_3$. This can be fixed by
adding new superfields charged under $SU(3)_c$ but singlet under the
other gauge subgroups, as pointed out in \cite{Majee:2007uv,Borah:2010kk}. 
Two examples are shown in Figure~\ref{fig:gcu-solved}, where the addition
of triplets of $SU(3)_c$ has been considered. To the left, one
generation is added at $m_{SUSY}$, whereas to the right five
generations are added at $v_{BL}$. In both cases the new contributions
to the running of $g_3$ are sufficient to obtain gauge coupling
unification.

However, this picture might be a bit too simple: the authors of 
reference \cite{Majee:2007uv} pointed out that thresholds effects 
at the GUT scale can lead to important
corrections to all gauge couplings. Note that these effects cannot
be calculated unless a  complete GUT model is specified and 
its high energy spectrum found.  Since our model is motivated by an underlying $SO(10)$ gauge
it is necessary to embedd also the $\Delta$ and $\Omega$ fields in complete 
representations. This will also lead most likely to a shift of $g_2$ and $g_{BL}$ and the condition
$g_2 = g_{BL}$ using just the running values might lead a wrong 
prediction for the GUT scale. 

We fix this problem by using a fixed
GUT scale $m_{GUT}=2 \cdot 10^{16}$ GeV throughout the
paper\footnote{Note, that when using 2-loop RGEs one does in general not have
strict unification \cite{Weinberg:1980wa,Ovrut:1980uv,Hall:1980kf}
and one has to define a scale where the threshold corrections
due to heavy degree of freedom have to be calculated, which is
perfectly consistent with out approach.}. 
In general, the three gauge couplings are different at this scale, but as
long as $v_{BL}$ and $v_R$ are not much below $m_{GUT}$ the differences are not
very large, e.g.\ of the order as in the usual MSSM. These differences would have 
important consequences in the  gauge couplings and spoil known relations 
normally obtained by using unified gaugino masses. To take this into
account we use gaugino masses at the chosen boundary scale of  $2 \cdot 10^{16}$ GeV
which will unify at the correct GUT scale as explained in sec.~\ref{sec:spectrum}.
Another advantage of this
approach is that our GUT scale is no longer sensitive to the high
energy VEVs $v_{BL}$ and $v_R$. This has an important consequence: if we would 
have applied naively the universal SUSY breaking boundary conditions at $\tilde{m}_{GUT}$, a large
theoretical uncertainty in the determination of the low energy parameters would
be present in our numerical conclusions.

Let us now discuss another important theoretical issue: $U(1)$ mixing. In the numerical
implementation of the model we have chosen to break the left-right symmetric gauge sector down to the SM
gauge group in one step in contrast to the heavy fields which we integrate out at two different scales. 
It would have been also possible to assume a gauge symmetry breaking in two
steps, i.e. $SU(2)_R \times U(1)_{B-L} \rightarrow U(1)_R \times U(1)_{B-L}
\rightarrow U(1)_Y$. 
However, the co-existence of two Abelian gauge groups
introduces the possibility of kinetic mixing, because a term of the form $\kappa
F_{\mu \nu}^R F^{B-L, \mu \nu}$ is allowed by gauge invariance
\cite{Holdom:1985ag}. Even if we assume that this term were absent at the
$SU(2)_R$ breaking scale it would be introduced by RGE
running already at 1-loop
level because the two $U(1)$'s are not orthogonal for the given particle
content. This might be indeed relevant because it has been shown that the
effect of kinetic mixing can be sizable even if the scale where the $U(1)$'s
coexist is rather short \cite{Malinsky:2005bi}. 
Using the procedure presented in \cite{Fonseca:2011vn} we have
checked that the use of a two step breaking and running the parameters
in the $U(1)_R \times U(1)_{B-L}$ basis is equivalent to the
one step breaking and running the parameters under the resulting
$U(1)_Y$. Here, we had to use 
1-loop boundary conditions and threshold effects to compensate
 differences in the 2-loop running as we use 2-loop RGEs.
  Nevertheless, although
kinetic mixing is conceptually interesting, in the model under
consideration its effects on the phenomenology are minor.

\subsection{Low energy spectrum} \label{sec:spectrum}

The introduction of additional superfields with masses below 
the GUT scale changes the RGEs not only for the gauge couplings 
but also for all MSSM soft terms. This in turn leads to changes 
in the electroweak scale supersymmetric mass spectra with 
respect to the standard CMSSM expectations. The CMSSM is defined
by the following  parameters: a common scalar mass $m_0$, 
a common gaugino mass parameter $M_{1/2}$ and common trilinear
parameter $A_0$ which are specified at the GUT scale. In addition
$\tan\beta$ is specified at the electroweak scale and the sign
of $\mu$ is fixed. As discussed in section \ref{subsec:gauge}
we do not have exact unification of the gauge couplings but 
expect that threshold corrections can account for the difference.
Also the gaugino mass parameters are subject to corrections of the
same size. Therefore we define the boundary conditions for the
gaugino mass parameters at the GUT scale $m_{GUT}$ as follows:
\begin{equation}
M_i = \frac{\alpha_i}{\alpha_{BL}} M_{1/2}
\end{equation}
For the calculation of the parameters at the SUSY scale
 we have used the complete 2-loop
RGEs at all scales which have been derived with the Mathematica package {\tt SARAH}
\cite{Staub:2008uz,Staub:2009bi,Staub:2010jh}. The numerical evaluation of the
RGEs as well as the calculation of the loop corrected masses was done with
{\tt SPheno} \cite{Porod:2003um,Porod:2011nf} using the {\tt SPheno} interface of {\tt SARAH} \cite{Staub:2011dp}. 
For more details about the numerical
implementation see also \cite{Esteves:2010si}.

\begin{figure}[t]
\begin{center}
\vspace{5mm}
\includegraphics[width=0.7\textwidth]{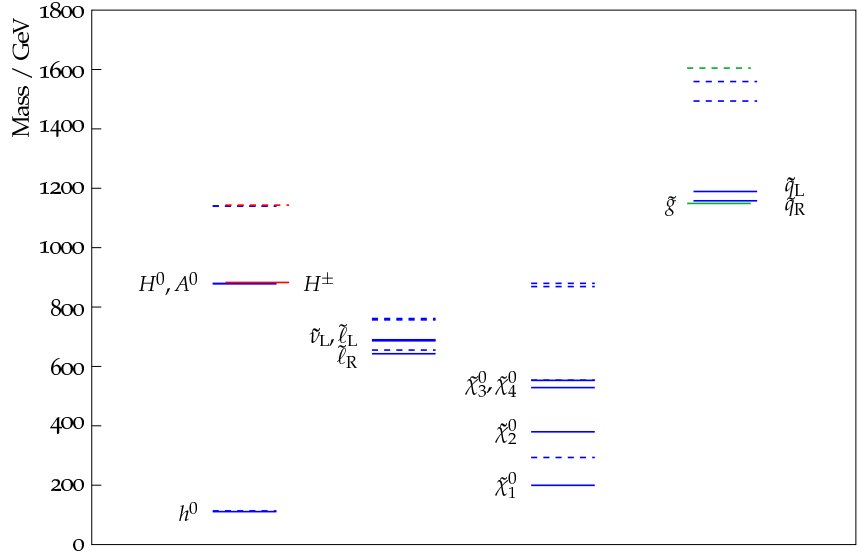}
\end{center}
\vspace{-5mm}
\caption{Example spectra comparing CMSSM with the $\Omega$LR model. 
Parameters have been chosen as $m_0=600$ GeV, $M_{1/2}=700$ 
GeV, $A_0=0$, $\tan\beta=10$, $\mu > 0$ and  $v_R =v_{BL}=10^{14}$ GeV. 
Full lines: $\Omega$LR, dashed line pure CMSSM. Note that in this example one has $m(\tilde{\chi}_2^0)_{CMSSM} \simeq m(\tilde{\chi}_4^0)_{\Omega LR}$ by accident.}
\label{fig:spec}
\end{figure}

An example spectrum is 
given in Figure~\ref{fig:spec}
where we show two SUSY (and Higgs spectra) for a 
specific choice of CMSSM parameters ($m_0=600$ GeV, $M_{1/2}=700$ 
GeV, $A_0=0$, $\tan\beta=10$ and $\mu > 0$). Dashed lines 
are the mass spectra for the pure CMSSM case, while the 
full lines have been calculated for the $\Omega$LR model 
with the specific choice of $v_R =v_{BL}=10^{14}$ GeV. These 
parameters were chosen to lie outside the region already excluded 
by the SUSY searches at the LHC experiments
\cite{Khachatryan:2011tk,Aad:2011hh,daCosta:2011qk,Chatrchyan:2011zy}, 
but are 
otherwise arbitrary. These spectra should not be taken as 
predictions for the $\Omega$LR model, but serve only for the 
sake of discussing the main differences between this model and 
the pure CMSSM case. 
In general, with $v_{BL}\le v_R < m_{GUT}$, the $\Omega$LR has a 
lighter spectrum than CMSSM for the same parameters. One 
can understand this semi-quantitatively with the help of the following 
considerations. 

Gaugino mass parameters run like gauge couplings do. Since in the 
$\Omega$LR model the running of the gauge couplings is changed with 
respect to the MSSM case, also gaugino masses are changed. Consider 
for example $M_1$. At 1-loop leading-log order we find 
\begin{eqnarray}\label{eq:M1}
M_1(m_{SUSY}) &=& M_{1/2} \:
\frac{\alpha_Y(m_{SUSY})}{\alpha_{BL}(m_{GUT})}  \\ &=&
M_{1/2} \: \left[ X_1 + X_2 \left( -3 l_1 + l_2 \right) \right]
\nonumber
\end{eqnarray}
where we have defined
\begin{eqnarray}
X_1 &=& \frac{5}{2} - \frac{3}{2}
\frac{\alpha_Y(m_{SUSY})}{\alpha_2(m_{SUSY})} - \frac{15
\alpha_Y(m_{SUSY})}{2 \pi} \ln \left(\frac{m_{GUT}}{m_{SUSY}}\right)
\, ,\, X_2 = \frac{3 \alpha_Y(m_{SUSY})}{2 \pi} 
\end{eqnarray}
and 
\begin{eqnarray}
l_1 &=& \ln \left(\frac{m_{GUT}}{v_R}\right) 
\, ,\,l_2 = \ln \left(\frac{v_R}{v_{BL}}\right) .
\end{eqnarray}
Note that, due to their definition, $l_{1,2} \ge 0$, with $l_1 = l_2 =
0$ in the CMSSM limit. One can easily check that $X_1 > 0$ and $X_2 > 0$.
For $m_{GUT} = 2 \cdot 10^{16}$ GeV and $m_{SUSY} = 1$ TeV one
obtains roughly $X_1 \simeq 0.424$ and $X_2 \simeq 0.008$. Therefore,
$M_1(m_{SUSY})$ decreases with $l_1$ and increases with $l_2$. In
other words, one can decrease $M_1(m_{SUSY})$ with respect to the
CMSSM value by using a large $l_1$ and $l_2 = 0$ and increase it 
only in the case $l_2 > 3  l_1$.  Similar equations 
hold for $M_2$ and $M_3$. However, at low energy ratios such as 
$M_1/M_2$ still follow the standard CMSSM expectations, only the 
relation to $M_{1/2}$ is changed. 

Similarly, sfermion mass parameters can be written schematically 
as, $m_{\tilde f}^2 \simeq m_0^2 + c_{\tilde f} M_{1/2}^2$, where the 
coefficients $c_{\tilde f}$ are different for different sfermions 
and depend on $v_R$ and $v_{BL}$. Exact expressions are given 
in appendix \ref{ap:inv}. One can show that for $v_{BL}\le v_R < m_{GUT}$
the $c_{\tilde f}$ are always smaller than in the CMSSM limit, explaining 
why also the sfermions in the $\Omega$LR model are lighter than 
in CMSSM, see Figure~\ref{fig:spec}. It is important to note, that 
gaugino masses change faster with $v_R$ than sfermion masses. 
We will come back to this point in the discussion about 
dark matter, see section~\ref{sec:dm}.

\begin{figure}
\begin{center}
\vspace{5mm}
\includegraphics[width=0.6\textwidth]{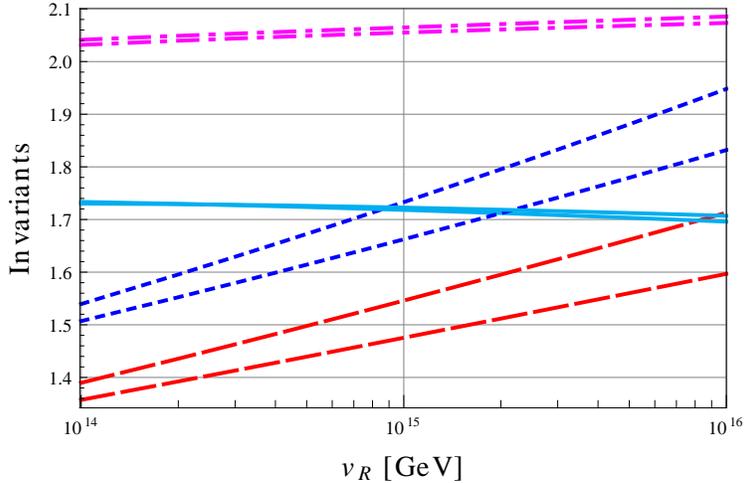}
\end{center}
\vspace{-5mm}
\caption{Invariant mass combinations QE/10 (purple, dashed-dotted
lines), LE (red, dashed lines), QU (dark blue, dotted lines) and
DL/10 (light blue, solid lines) as functions of $v_R$. $v_{BL} \in
[10^{12},10^{14}]$ GeV, upper lines correspond to $v_{BL} = 10^{12}$
GeV and lower lines to $v_{BL} = 10^{14}$ GeV.}
\label{fig:invariants}
\end{figure}

Individual SUSY masses depend strongly on the initial values of 
$m_0$ and $M_{1/2}$.  However, one can form four different combinations 
(``invariants'') of soft SUSY breaking parameters, which we choose as 
\begin{eqnarray}\label{eq:4Inv}
\begin{array}{c}
LE \equiv (m_{L}^2 -m_{e^c}^2)/M_1^2, \\
QE \equiv (m_{Q}^2 -m_{e^c}^2)/M_1^2, \\
DL \equiv (m_{d^c}^2 -m_{L}^2)/M_1^2, \\
QU \equiv (m_{Q}^2 -m_{u^c}^2)/M_1^2, 
\end{array}
\end{eqnarray}
where at the leading-log level $m_0$ and $M_{1/2}$ drop out. 
Formulas for an analytical 1-loop leading-log calculation of 
these four invariants are given in appendix \ref{ap:inv}.

Figure~\ref{fig:invariants} shows the analytically calculated values 
of the invariants as a function of $v_R$ for two values of $v_{BL}$. 
The invariants depend more strongly on $v_R$ and only weakly on $v_{BL}$. 
As the figure shows, at leading order QE and DL show a very mild
dependence 
on the scales $v_R$ and $v_{BL}$, while LE and QU {\em decrease} 
with decreasing $v_R$. At the $v_R$ scale one has $m_{L}^2 =
m_{e^c}^2$ 
and $m_{Q}^2 =m_{u^c}^2 = m_{d^c}^2$ due to parity
conservation. This implies that 
the lower $v_R$ is, the smaller the difference between them at the SUSY scale
is, which means that the $LE$ and $QU$ invariants can have values
below the CMSSM prediction. We have checked that this effect is quite
robust, and the theoretical uncertainties do not have much influence
on it. In particular, these two invariants do not show any
numerical dependence on $m_{GUT}$, due to cancellations among left and
right contributions in the running from the GUT scale to $v_R$. 
This is an interesting result, since in 
CMSSM models with a type-II or type-III seesaw, all four invariants 
are always {\em larger} than their CMSSM limit for seesaw scales 
below the GUT scale \cite{Buckley:2006nv,Hirsch:2008gh,Esteves:2010ff}. 
This allows, at least in principle, to distinguish between models with 
a high scale LR group and CMSSM models with type-II or type-III 
seesaw as the explanation for the observed neutrino masses.

\begin{figure}
\begin{center}
\vspace{5mm}
\includegraphics[width=0.49\textwidth]{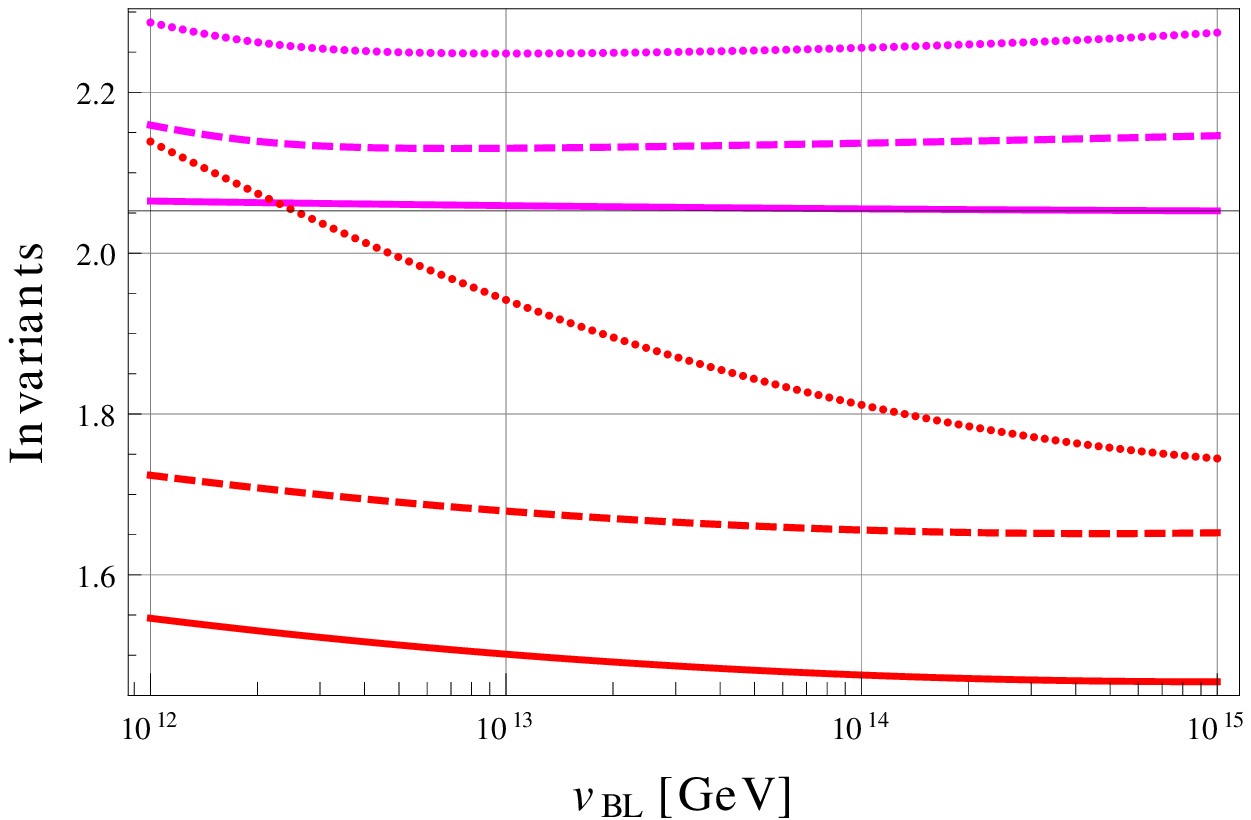}
\includegraphics[width=0.49\textwidth]{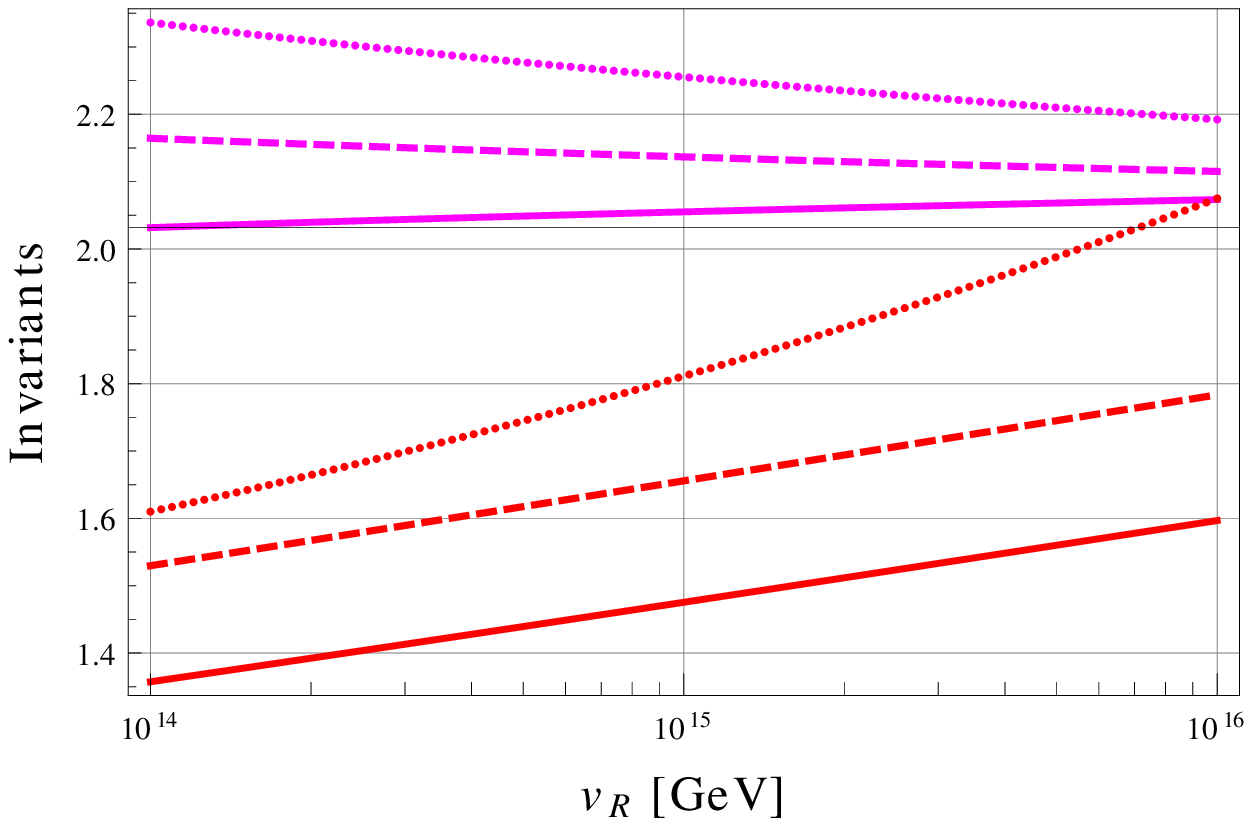}
\end{center}
\vspace{-5mm}
\caption{Invariant mass combinations QE/10 and LE as functions
of $v_{BL}$ (left) and $v_R$ (right). The color code is the same as
in Figure~\ref{fig:invariants}. Solid lines correspond to the
analytical approximation, dashed lines to 1-loop numerical results and
dotted lines to 2-loop numerical results. Similar deviations are found for the
other two invariant mass combinations.}
\label{fig:invariantscomp}
\end{figure}

Figure~\ref{fig:invariants} shows the leading-log calculation. It is 
known in the case of seesaw based models, that there exist important 
2-loop corrections and 1-loop threshold effects 
\cite{Hirsch:2008gh,Esteves:2010ff}. Similarly, also in the 
$\Omega$LR model important numerical corrections exist, as is 
shown in Figure~\ref{fig:invariantscomp}. When the invariants are 
written including the effect of non-unification at the  GUT scale, 
see appendix \ref{ap:inv}, they reproduce qualitatively the numerical results. 
However, as shown in Figure~\ref{fig:invariantscomp} quantitatively 
important shifts are obtained going from leading-log to (numerically 
solved) full 1-loop calculation. Even going from 1-loop to 2-loop 
calculation, numerically important differences are found. 

To summarize, the invariants are good model discriminators in principle. 
Especially noteworthy,  in the $\Omega$LR model LE and QU are expected 
to be below their CMSSM limit. However, to identify such spectrum 
distortions, once the  SUSY mass spectrum is measured, 
will require highly accurate measurements and therefore 
measurements at an ILC.

\subsection{Lepton flavour violation and the role of $\theta_{1,2}$} \label{sec:low-energy}

Flavour violation in leptonic processes has attracted a lot of attention in the
experimental community. Decays like $\mu \to e \gamma$ have been searched for
decades, without any positive result. Very recently, the MEG experiment
\cite{meg} has published the results of the analysis of the data collected in
2009 and 2010 \cite{Adam:2011ch}, setting the new bound $\text{Br}(\mu \to e
\gamma) < 2.4 \cdot 10^{-12}$. This impressive experimental limit strongly
constrains models with extended lepton sectors, such as the $\Omega$LR 
model.

The branching ratio for $l_i \to l_j \gamma$ can be written as
\cite{Kuno:1999jp}
\begin{equation} \label{brLLG}
\text{Br}(l_i \to l_j \gamma) = \frac{48 \pi^3 \alpha}{G_F^2} \left(
|A_L^{ij}|^2 + |A_R^{ij}|^2 \right) \text{Br}(l_i \to l_j \nu_i \bar{\nu}_j)
\thickspace .
\end{equation}
The couplings $A_L$ and $A_R$ are generated at the 1-loop level. 
The relation between these couplings
and the slepton soft masses is given approximately by
\begin{equation} \label{A-dependence}
A_L^{ij} \sim \frac{(m_L^2)_{ij}}{m_{SUSY}^4} \quad , \quad A_R^{ij} \sim \frac{(m_{e^c}^2)_{ij}}{m_{SUSY}^4} \thickspace, 
\end{equation}
where $m_{SUSY}$ is a typical supersymmetric mass. In the derivation of this estimate it has been 
assumed that (a) chargino/neutralino masses are similar to slepton 
masses and (b) A-terms mixing left-right transitions are negligible.

Note that, due to the negligible off-diagonal entries in 
$m_{e^c}^2$, a pure seesaw model predicts $A_R \simeq 0$. On the contrary, in
reference \cite{Esteves:2010si} it was pointed out that a left-right symmetry at
high energies induces non-negligible off-diagonal elements in $m_{e^c}^2$,
giving additional contributions to LFV processes. In fact, taking into account the running from the GUT scale to the $v_R$ 
scale, the off-diagonal elements of the slepton soft masses at 1-loop order can be written in leading-log approximation as \cite{Chao:2007ye,Esteves:2010si}
\begin{eqnarray}
\left( m_L^2 \right)_{ij}|_{v_R} &=& - \frac{1}{4 \pi^2} \left( 3 f f^\dagger + \sum_{k=1}^2 Y_L^{(k)} Y_L^{(k) 
\: \dagger} \right)_{ij} (3 m_0^2 + A_0^2) \ln \left( \frac{m_{GUT}}{v_R} \right) \thickspace,
\label{apprge1} \\
\left( m_{e^c}^2 \right)_{ij}|_{v_R} = \left( m_{L^c}^2 \right)_{ij}|_{v_R} &=& - \frac{1}{4 \pi^2} \left( 3 f^\dagger f + \sum_{k=1}^2 Y_L^{(k) 
\: \dagger} Y_L^{(k)} \right)_{ij} (3 m_0^2 + A_0^2)
\ln \left( \frac{m_{GUT}}{v_R} \right) \thickspace, \label{apprge2}
\end{eqnarray}
which are of the same size in the CP conserving case.  
The $A$ parameters also develop LFV off-diagonals in the 
running. We do not give the corresponding approximated equations 
because they do not lead to qualitatively new features.

The angular distribution of the outgoing positron at, 
for example, the MEG experiment could be used to discriminate between 
left- and right-handed polarized states \cite{Okada:1999zk,Hisano:2009ae}. 
If MEG is able to measure the positron polarization asymmetry, defined as
\begin{equation}
A_{LR} = \frac{|A_L|^2-|A_R|^2}{|A_L|^2+|A_R|^2},
\end{equation}
we will have an additional tool to distinguish from pure seesaw models, where $A_{LR} \simeq 1$ is predicted.

We extend here the discussion of ref.~\cite{Esteves:2010si}, 
addressing the influence of the high energy
parameters on low energy observables. As explained in section \ref{sec:model},
the mixing angles of the bidoublets at the $v_R$ scale have a very strong impact
on low energy LFV processes. As is
shown in eqs.~\eqref{matching-1} and
\eqref{matching-2}, this is due to their relation to the Yukawa parameters. If
the bidoublet couplings are tuned such that $\theta_1$ and $\theta_2$
get very close values, the entries of the Yukawa matrix $Y_L$  will become very
large, which in turn will lead to large running effects for the soft squared
masses of the sleptons. Therefore, we expect a strong correlation between
$s_{21} = \sin(\theta_2 - \theta_1)$ 
and the size of the LFV processes.

\begin{figure}
\begin{center}
\vspace{5mm}
\includegraphics[width=0.49\textwidth]{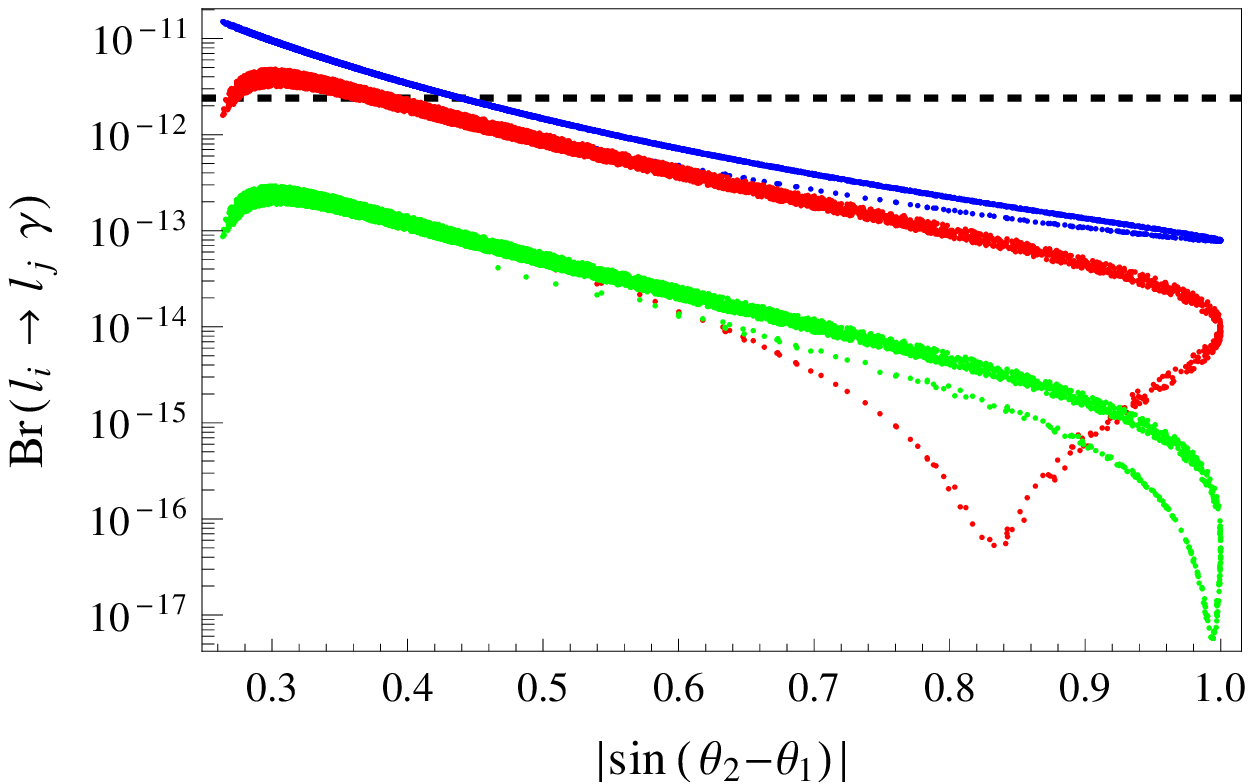}
\includegraphics[width=0.49\textwidth]{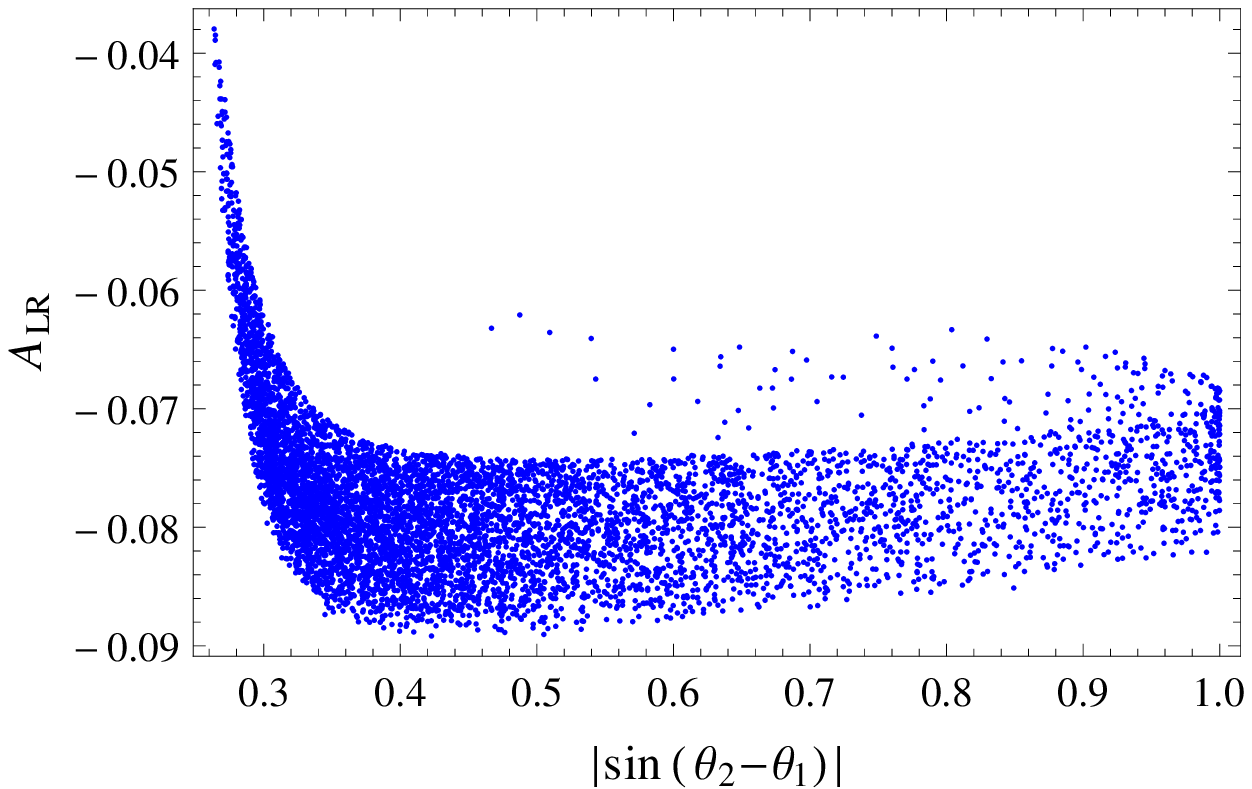}
\end{center}
\vspace{-5mm}
\caption{To the left, $\text{Br}(\mu \to e \gamma)$ (blue),
$\text{Br}(\tau \to \mu \gamma)$ (red) and $\text{Br}(\tau \to e
\gamma)$ (green) as a function of $|\sin(\theta_2 - \theta_1)|$. The
horizontal line shows the current MEG bound $\text{Br}(\mu \to e
\gamma) < 2.4 \cdot 10^{-12}$.  To the right, the positron
polarization asymmetry $A_{LR}$ as a function of $|\sin(\theta_2 -
\theta_1)|$. Both figures have been made fitting neutrino data with
the flavour structure in $Y_\nu$ and taking the parameter choice $M_S
= 10^{12}$ GeV, $v_{BL} = v_R = 10^{15}$ GeV for the CMSSM point $m_0
= 120$ GeV, $M_{1/2} = 600$ GeV, $\tan \beta = 10$, $A_0 = 0$ and $\mu
> 0$.}
\label{fig:randtheta}
\end{figure}

The left side of Figure~\ref{fig:randtheta} shows such correlation for a
particular but representative set of parameters. We note that a proper choice of
$\theta_2 - \theta_1$ can enhance (or suppress) the $l_i \to l_j \gamma$
branching ratios by several orders of magnitude. Moreover, special values for
$\theta_2 - \theta_1$ are found where the $\tau$ decays have strong
cancellations. Their origin can be traced back to the flavour structure of the
LFV entries in $m_L^2$ and $m_{e^c}^2$.

The RGE running from $m_{GUT}$ to $v_R$ introduces LFV entries in $m_L^2$ and
$m_{e^c}^2$ proportional to $\sum_k Y_L^{(k)} Y_L^{(k) \: \dagger}$ (the dagger is to
be exchanged for the case of $m_{e^c}^2$, which makes no difference if CP is
conserved). Expanding this expression in terms of
$Y_e$ and $Y_\nu$, the Yukawa couplings at $v_R$, one obtains
\begin{equation}
\sum_k Y_L^{(k)} Y_L^{(k) \: \dagger} = \frac{1}{s^2_{21}} \left[ Y_\nu Y_\nu^\dagger
+ Y_e Y_e^\dagger + c_{21} ( Y_e Y_\nu^\dagger + Y_\nu Y_e^\dagger ) \right]
\end{equation}
where $c_{21} = \cos(\theta_2 - \theta_1)$. For the observable $\text{Br}(l_i
\to l_j \gamma)$ we have to compute $( \sum_k Y_L^{(k)} Y_L^{(k) \: \dagger} )_{ij}$,
with $i \ne j$. The $Y_e Y_e$ piece can be neglected, since $Y_e$ is almost
diagonal (we choose to work in the basis in which $Y_e$ is diagonal at the
electroweak scale). Furthermore, in the approximation $m_{l_i} \gg m_{l_j}$ one
can also neglect the $Y_\nu Y_e^\dagger$ piece, since it will be proportional to
$m_{l_j}$. Therefore, using eqs.~\eqref{brLLG} and \eqref{A-dependence}, one
finds
\begin{equation} \label{brLLG-2}
\text{Br}(l_i \to l_j \gamma) \propto \frac{1}{s^4_{21}} | ( Y_\nu Y_\nu^\dagger + c_{21} Y_e Y_\nu^\dagger )_{ij} |^2
\end{equation}
From eq.~\eqref{brLLG-2} one can see that as $\theta_1$ and $\theta_2$ get
closer $s_{21} \to 0$ and one gets an enhancement.
The interesting new point
is the interplay of the two terms as for specific values of
$\theta_2 - \theta_1$  cancellations occur. In case of
of $\text{Br}(\mu \to e \gamma)$ this hardly occurs
as in general  the $\mu$-Yukawa coupling is much smaller
than the $|(Y_\nu)_{ij}|$ and a cancellation is not possible.
On the contrary, the $\tau$-Yukawa coupling
is comparable to $Y_\nu$ in large parts of interesting
parameter space, allowing for
potential cancellations.

However, we want to stress that the position and the degree of these
cancellations depend on the high energy VEVs $v_{BL}$ and $v_R$ and 
thus one
cannot determine $\theta_2 - \theta_1$ uniquely. Moreover, this
connection between leptons and bidoublet mixing angles is lost if one fits
neutrino data using the superpotential coupling $f$, which gives rise to the
Majorana mass of the right-handed neutrino. In this possibility, called '$f$ fit' in
\cite{Esteves:2010si}, there is no dependence on the bidoublet mixing angles,
since $f$ couples the leptons to the $\Delta$ triplets, but not to the
bidoublets. 
For completeness we note that also the positron polarization
asymmetry $A_{LR}$ depends somewhat on $\theta_2 - \theta_1$ as can
be seen on the right side of Figure~\ref{fig:randtheta}. 

Finally, we have  checked that this strong dependence on $\theta_{1,2}$ is not
present in the quark sector. Flavour violating processes, such as $b\to
s\gamma$, are very weakly affected by the bidoublet mixing angles and the
phenomenology follows very closely the well-known results of the CMSSM. This is
due to the fact that flavour violation in the (s)quark sector is dominated
by the CKM matrix. In other words, the $\Omega$LR model is a minimal
flavour violating one \cite{Buras:2000dm,DAmbrosio:2002ex}.

\section{Dark matter}
\label{sec:dm}

Astrophysical observations and the data from WMAP \cite{Komatsu:2010fb} 
put on solid grounds the existence of non-baryonic dark matter in the 
universe. The PDG \cite{Nakamura:2010zzi} quotes the value 
$\Omega_{\rm DM}h^2=0.110\pm 0.006$ at $1\sigma$ C.L. It is well-known 
that with this low value of $\Omega_{\rm DM}h^2$ only a few, very 
specific regions in the parameter space of CMSSM can give the 
correct relic density for the neutralino \cite{Griest:1990kh}. 
These are known as (in no specific order) (i) the stau co-annihilation 
region \cite{Ellis:1998kh}; (ii) the stop co-annihilation region 
\cite{Boehm:1999bj,Ellis:2001nx,Edsjo:2003us}; (iii) the focus point 
line \cite{Feng:1999zg,Feng:2000gh} and (iv) the Higgs funnel 
\cite{Griest:1990kh}. 

In any of the co-annihilation regions the relic density of the 
lightest neutralino, which in the above cases (i) and (ii) is 
mostly bino, is reduced with respect to naive expectations by a 
close mass degeneracy between the neutralino and the NLSP. Any 
changes in the SUSY spectra can then have a rather large impact 
on the calculated $\Omega_{\tilde\chi^0_1}h^2$. In the focus point 
region a comparatively small value of $\mu$, with respect to 
the rest of the CMSSM parameter space, leads to an increased 
higgsino content in the lightest neutralino. This in turn leads 
to a larger coupling of the LSP to the $Z^0$, thus reducing 
the relic density. The fourth allowed region, the Higgs funnel,
appears for high values of $\tan\beta$ and for parameters where 
the CP-odd Higgs scalar $A$ has a mass which is 
twice the LSP mass. In this case there is an s-channel resonant 
enhancement of the neutralino annihilation cross section and 
large values of $\tan\beta$ are needed for this resonance to 
be effective, since the width of $A$ is large at large $\tan\beta$. 
All of the regions discussed above have been extensively studied in 
the CMSSM \cite{Ellis:1983ew,Griest:1990kh,Drees:1992am,Jungman:1995df}
and in some extensions of it, like CMSSM plus seesaw of either 
type-II or type-III \cite{Esteves:2009qr,Esteves:2010ff}. 

Below we discuss the changes for these allowed regions for the
$\Omega$LR model. We have used {\tt SPheno} to compute the low energy
spectrum and used it as an input for {\tt micrOmegas}
\cite{Belanger:2006is} to obtain the value for
$\Omega_{\tilde\chi^0_1}h^2$. To include effects of flavour violation,
we have created suitable model files for {\tt micrOmegas} with {\tt
  SARAH}.  In the present setup the main effect on dark matter is via
changes in spectrum (see section \ref{sec:spectrum}).  For low
$v_{BL}$ or $v_R$ the dark matter allowed regions are reduced compared
to the CMSSM.  We have found this effect for the stau co-annihilation,
stop co-annihilation, Higgs funnel and focus point regions and discuss
these regions in turn.

\subsection{Non-flavoured dark matter allowed regions in $\Omega$LR}

\begin{figure}
\begin{center}
\vspace{5mm}
\includegraphics[width=0.47\textwidth]{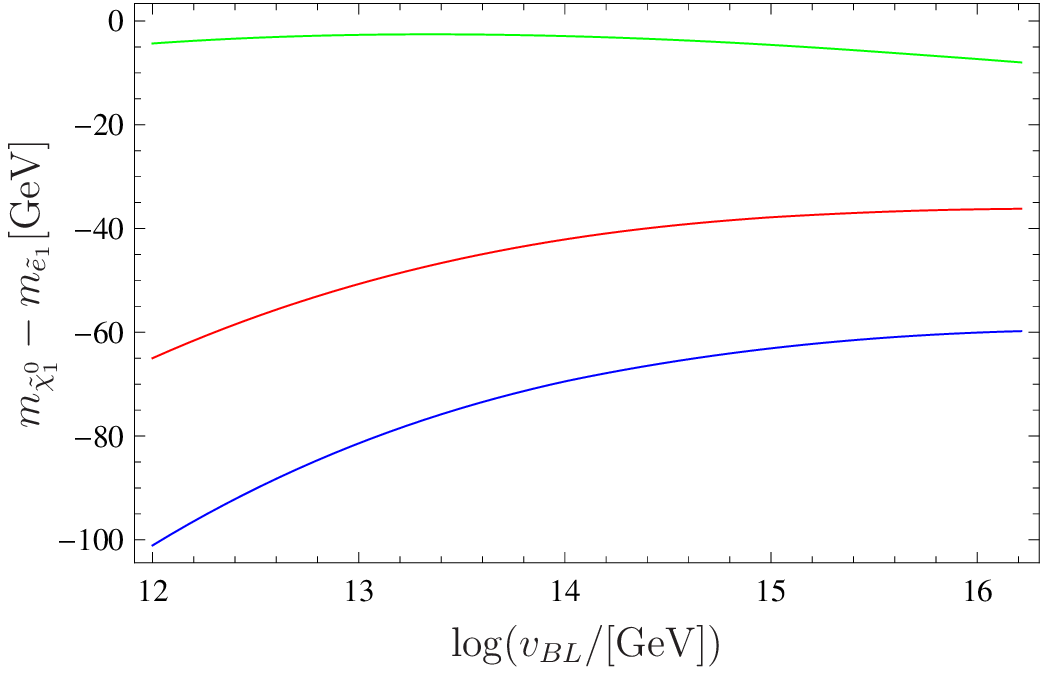}
\hspace{5mm}
\includegraphics[width=0.47\textwidth]{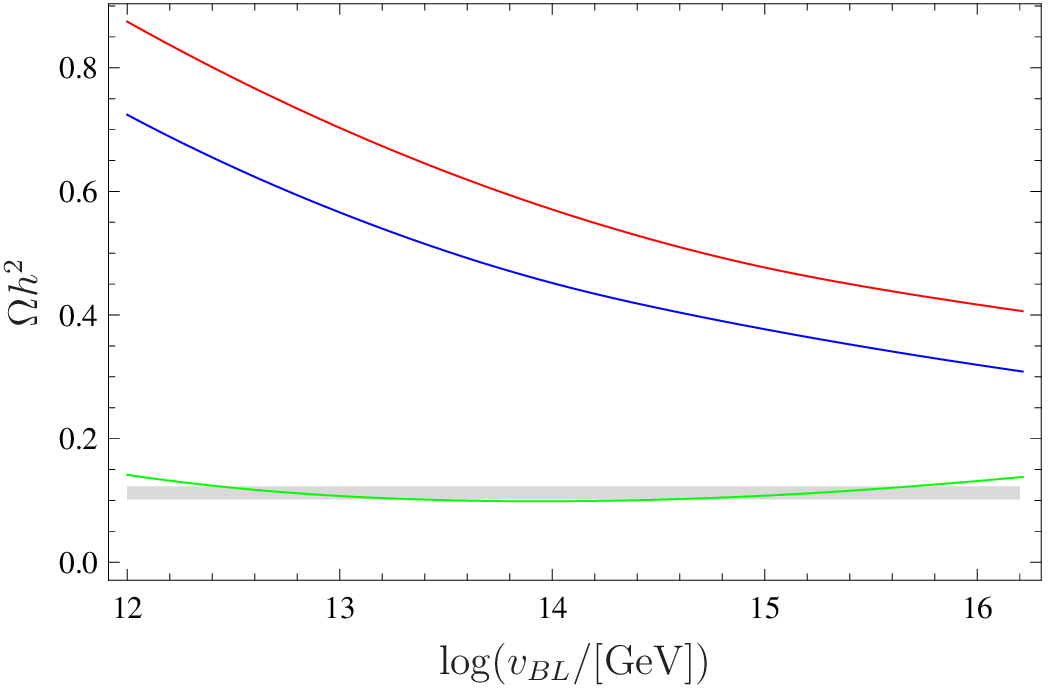}
\end{center}
\vspace{-5mm}
\caption{Stau co-annihilation. On the left: mass difference between the lightest
neutralino and the lightest stau vs. $v_{BL}$ for three different values of
$v_R = 10^{12}$ (blue), $10^{14}$ (red) and $10^{16}$~GeV (green). On the right:
resulting dark matter relic density $\Omega h^2$. The gray band shows the
$\Omega h^2 = [0.1018,0.1228]$. The other input parameters have been chosen 
as: $m_0=105$~GeV, $M_{1/2}=600$~GeV, $\tan\beta=10$, $A_0=0$~GeV.}
\label{fig:staucoa}
\end{figure}

\begin{figure}
\begin{center}
\vspace{5mm}
\includegraphics[width=0.47\textwidth]{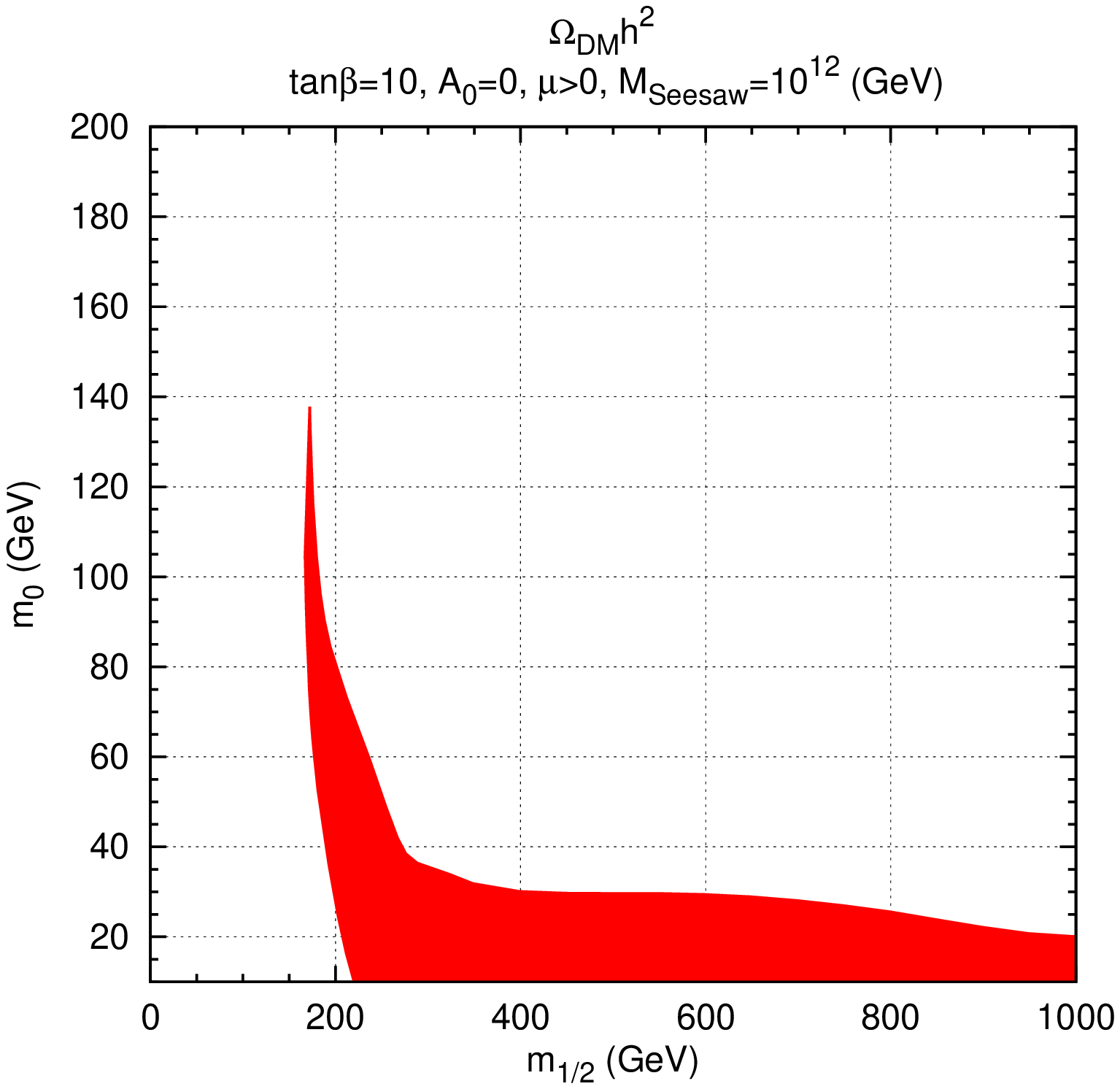}
\hspace{5mm}
\includegraphics[width=0.47\textwidth]{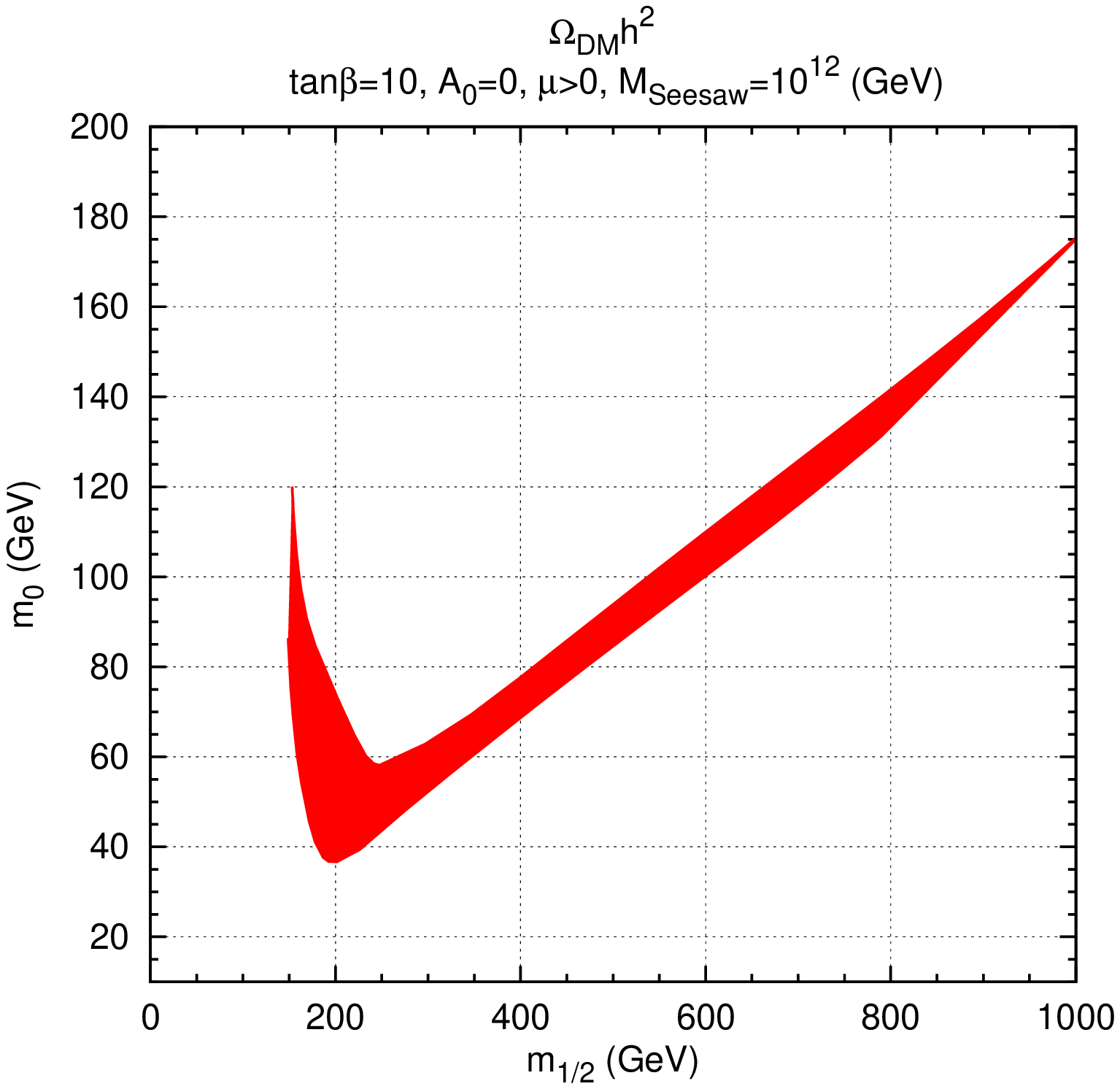}
\caption{Dark Matter allowed regions for $\tan\beta=10$, and
  $v_{BL}=v_R=1.5\times10^{15}$ GeV (left) and $v_{BL}=v_R=10^{16}$
  GeV (right). For this figure we have chosen a low seesaw scale, $M_S
  = 10^{12}$ GeV, which assures that the LFV constraints are
  respected.}
\label{fig:DMtb10-I}
\end{center}
\end{figure}

\subsubsection{Stau co-annihilation}

As discussed in section \ref{sec:spectrum}, lowering the values of 
$v_{BL}$ and $v_R$ leads in general to a lighter spectrum compared to 
the pure CMSSM case, for the same values of CMSSM parameters. In 
Figure~\ref{fig:staucoa} we show an example for the mass difference 
between the lightest neutralino and the lightest scalar tau for three 
values of $v_R$ as a function of $v_{BL}$. Since the mass of the bino 
decreases faster than the mass of the stau, the mass difference 
increases and this in turn leads to a larger value of the neutralino 
density, as shown in the plot on the right, because co-annihilation 
becomes ineffective for $\Delta_m= m_{\tilde\tau_1}-m_{\chi^0_1}$ 
larger than a few GeV. Note that the CMSSM parameters in this example 
have been chosen to give approximately the correct relic density for 
$v_R\simeq 10^{16}$ GeV.

Two examples of allowed stau co-annihilation regions are then shown 
in Figure~\ref{fig:DMtb10-I}. The figure shows the 3 $\sigma$ 
allowed regions for $\tan\beta=10$, and $v_{BL}=v_R=1.5\times10^{15}$ GeV 
(left) and $v_{BL}=v_R=10^{16}$ GeV (right). Lowering $v_R$ (and $v_{BL}$) 
shifts the allowed region towards smaller values of $m_0$. Depending 
on the values of $\tan\beta$ and $A_0$ chosen, for values of roughly 
$v_{BL}=v_R \sim 10^{15}$ GeV the stau coannihilation line disappears 
completely. The observation of a SUSY particle spectrum consistent 
with the stau co-annihilation region could thus be turned into a 
lower limit on $v_{BL}$ and $v_R$ within CMSSM, at least in principle. 

However, as has been noted in section \ref{sec:spectrum}, the effect 
of $v_R$ on $M_1$ is stronger and inverse to the one of $v_{BL}$. We 
have checked that it is possible to obtain a stau-coannihilation region 
slightly above the CMSSM expected region (for any fixed $\tan\beta$ and $A_0$) 
for values of $v_R$ close to the GUT scale and $v_{BL}$ low. Thus, 
the stau coannihilation can give a lower limit on $v_R$ only as a 
function of the (assumed) value of $v_{BL}$.

\subsubsection{Stop co-annihilation}

\begin{figure}
\begin{center}
\vspace{5mm}
\includegraphics[width=0.5\textwidth]{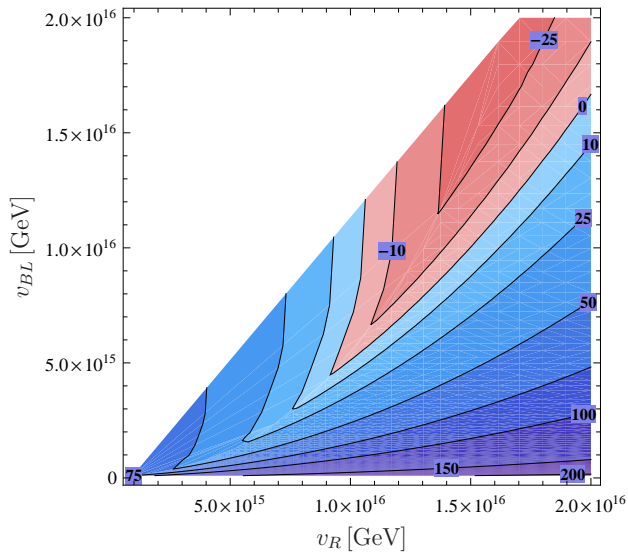}
\end{center}
\vspace{-5mm}
\caption{$m_{\tilde{\chi}_1^0} - m_{\tilde{t}_1}$ contours in the $v_{BL}-v_R$
plane. The CMSSM  parameters take the following values: $m_0 = 630$ GeV,
$M_{1/2}= 400$ GeV, $\tan\beta=10$, $A_0=-2$ TeV and $\mu>0$.}
\label{fig:stopcoa}
\end{figure}

With typical choices of CMSSM parameters the squarks are much heavier
then the LSP. However, if the off-diagonal elements in the squark mass
matrices are significant, the lighter mass eigenstates have their masses lowered 
and the lighter stop can be almost degenerate in mass with the lightest
neutralino. This stop co-annihilation has been studied in the literature
\cite{Boehm:1999bj,Ellis:2001nx,Edsjo:2003us} and happens for instance
if the soft breaking parameter $A_0$ has a large value  We
also explored this possibility in the present setup. Figure~\ref{fig:stopcoa}
shows an example. In this plot, the CMSSM parameters where chosen 
to be $m_0 = 630$ GeV, $M_{1/2}= 400$ GeV, $\tan\beta=10$,
$A_0=-2$ TeV and $\mu>0$, which leads to a mass difference between
stop and neutralino below 2 GeV in the pure CMSSM limit. Lowering 
$v_R$ or $v_{BL}$ below the GUT scale, this mass difference again 
increases, rendering co-annihilation ineffective. The lowest possible 
values for $v_{BL}$ and $v_R$ in this scenario are usually found 
in the region $v_{BL} \sim v_R$, as the figure shows. We found no 
combination of parameters which improves the result shown in the figure 
by more than a few GeV.  Our conclusion is therefore, that the stop 
co-annihilation region vanishes for low intermediate scales, similarly to 
the stau-coannihilation region. Observation of a SUSY spectrum consistent 
with stop co-annihilation could therefore be interpreted as a lower 
limit on $v_R$ and $v_{BL}$.

\subsubsection{Higgs funnel}

\begin{figure}
\begin{center}
\vspace{5mm}
\includegraphics[width=0.5\textwidth]{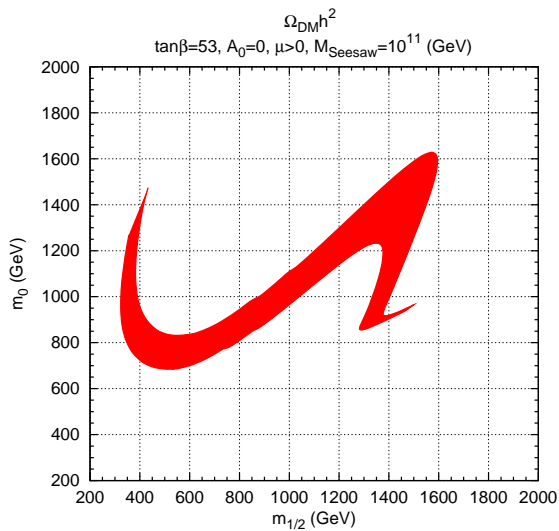}
\end{center}
\vspace{-5mm}
\caption{The allowed region for $\Omega_{\tilde\chi^0_1}h^2$ in the
  $m_0$-$M_{1/2}$ plane for $\tan \beta = 53$, $v_{BL}=v_R=10^{16}$
  GeV. For this figure we have chosen a low seesaw scale, $M_S =
  10^{11}$ GeV, which assures that the LFV constraints are respected.}
\label{fig:funnelexample}
\end{figure}

As in the previous two examples, we have found that also the 
Higgs funnel region depends rather strongly on the choice of 
$v_R$ and $v_{BL}$. As mentioned above, at large values of 
$\tan\beta$, typically $\tan\beta \ge 50$ the width of the 
CP-odd Higgs boson A can be large enough so that there is a 
s-channel resonance for the neutralino pair annihilation 
in the region $ 2 m_{\chi^0_1} \simeq m_A$. An example for this 
Higgs funnel region is shown in Figure~\ref{fig:funnelexample}. 

\begin{figure}
\begin{center}
\vspace{5mm}
\includegraphics[width=0.5\textwidth]{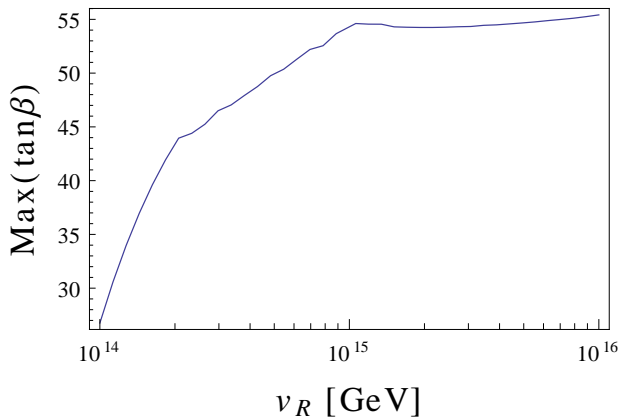}
\end{center}
\vspace{-5mm}
\caption{Upper limit on $\tan \beta$ as a function of $v_R$ for
$v_{BL} = 10^{14}$ GeV. The figure was obtained for the choice of
CMSSM parameters $m_0 = M_{1/2} = 1.5$ TeV, $A_0 = 0$ and $\mu > 0$.}
\label{fig:maxtanbeta}
\end{figure}

In the $\Omega$LR model the Higgs funnel region disappears when the
intermediate scales are lowered below a certain limit. This is due to
the existence of an upper limit on $\tan\beta$, stronger in the
$\Omega$LR model than in CMSSM, caused by the requirement of (1)
perturbativity of the bottom and tau Yukawa couplings, and (2)
stability of the electroweak symmetry breaking minimum of the Higgs
potential. In the CMSSM this limit is roughly $\tan \beta\simeq
55-60$.  In the $\Omega$LR model this limit depends on $v_R$, and
becomes stronger when this high energy scale is lowered. This can be
understood, in principle, from the fact that the lower $v_R$ is, the
larger the RGE effects on the Yukawa couplings are.  As shown in
Figure~\ref{fig:maxtanbeta}, which should be taken as an illustrative
example, for a given value of $v_R$ one can always find an upper limit
on $\tan \beta$, which in the limit $v_R$ approaching $m_{GUT} = 2
\cdot 10^{16}$ GeV recovers the usual CMSSM limit on $\tan \beta$. For
lower values of $v_R$, therefore, there is an upper limit on the A
width. The allowed Higgs funnel region becomes smaller until for a
certain value of $v_R$ it disappears completely. We have also observed
additional minor effects that can play a role in the determination of
the correct relic density. For example, a dependence of the higgsino
component of the LSP on $v_{BL}$ has been found. Again, this can lead
to the disappearance of the Higgs funnel.
Therefore, if SUSY is discovered with large $\tan\beta$, this could 
again be interpreted as a lower limit on $v_R$.

\subsubsection{Focus point}

The focus point is also affected by the running of the parameters above the parity breaking scale. The observed dark matter relic density is obtained in this region thanks to annihilation into $Z^0$ bosons, a process which is only effective when the higgsino component of the lightest neutralino is sufficiently large. This is provided by a small $\mu$ parameter, $\mu \sim M_1$.

As explained, $M_1(m_{SUSY})$ is typically smaller in the $\Omega$LR model than in the CMSSM. When one lowers the high energy VEVs $v_{BL}$ and/or $v_R$, the resulting $M_1(m_{SUSY})$ gets lowered as well, see eq. \eqref{eq:M1}. Therefore, the required tuning with the $\mu$ parameter can only be obtained by increasing $M_{1/2}$. A shift of the focus point region towards larger values of $M_{1/2}$ is thus expected.

\begin{figure}
\begin{center}
\vspace{5mm}
\includegraphics[width=0.5\textwidth]{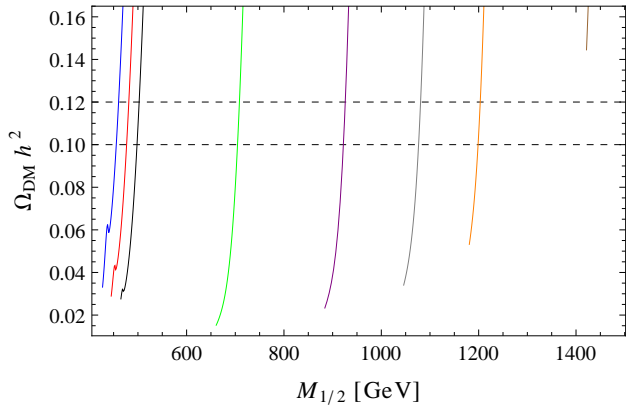}
\end{center}
\vspace{-5mm}
\caption{$\Omega_{DM} h^2$ as a function of $M_{1/2}$ for the choice of parameters $m_0 = 2.5$ TeV, $\tan\beta=30$, $A_0=0$ and $\mu>0$. From left to right the different curves correspond to $v_{BL} = v_R = [2 \cdot 10^{16}$, $1.9 \cdot 10^{16}$, $1.8 \cdot 10^{16}$, $10^{16}$, $5 \cdot 10^{15}$, $3 \cdot 10^{15}$, $2 \cdot 10^{15}$, $10^{15}]$ GeV. All curves reach their minimum, lower $M_{1/2}$ values would spoil electroweak symmetry breaking. The dashed horizontal lines show the values $\Omega_{DM} h^2 =
  [0.1018,0.1228]$.}
\label{fig:focus-1}
\end{figure}

This expectation has been check numerically. Figure \ref{fig:focus-1} shows $\Omega_{DM} h^2$ as a function of $M_{1/2}$ for the choice of parameters $m_0 = 2.5$ TeV, $\tan\beta=30$, $A_0=0$ and $\mu>0$. As $v_{BL}$ and $v_R$ are decreased, one needs to go to larger values of $M_{1/2}$ in order to reproduce the observed relic abundance. We find that for $v_{BL} = v_R \sim 10^{15}$ GeV one cannot make the annihilations sufficiently effective and the focus point disappears (for this particular choice of parameters). Note that in this figure all curves reach their minimum, lower $M_{1/2}$ values would spoil electroweak symmetry breaking. We also would like to emphasize that the dependence on $v_{BL}$ and $v_R$ is very strong. This can be clearly seen from the non-negligible shift in $M_{1/2}$ obtained when one goes from $v_{BL} = v_R = m_{GUT} = 2 \cdot 10^{16}$ GeV to just $v_{BL} = v_R = 1.9 \cdot 10^{16}$ GeV.

\begin{figure}
\begin{center}
\vspace{5mm}
\includegraphics[width=0.5\textwidth]{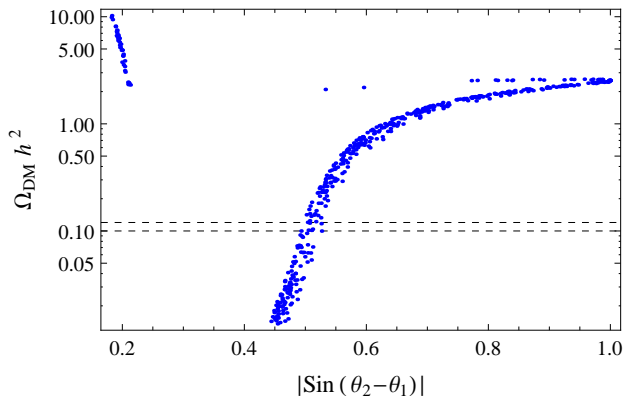}
\end{center}
\vspace{-5mm}
\caption{$\Omega_{DM} h^2$ as a function of $\sin (\theta_2 -
  \theta_1)$ for the choice of parameters $m_0 = 2.5$ TeV, $M_{1/2}=
  708$ GeV, $\tan\beta=30$, $A_0=0$, $\mu>0$ and $v_{BL} = v_R =
  10^{16}$ GeV. The dashed horizontal lines show the values
  $\Omega_{DM} h^2 = [0.1018,0.1228]$.}
\label{fig:focus-2}
\end{figure}

We also found a clear dependence of $\Omega_{DM} h^2$ on the angles $\theta_{1,2}$, as shown in figure \ref{fig:focus-2}. This can be easily understood from the matching conditions at the $SU(2)_R$ breaking scale, eqs. \eqref{matching-1} and \eqref{matching-2}. The running of the soft parameter $m_{H_u}^2$ is strongly affected by the quark Yukawas $Y_Q^{1,2}$ at energies above $v_R$. This leads to different values for the $\mu$ parameter at the SUSY scale, affects the annihilation cross-section of the lightest neutralino and modifies its relic abundance. Another relevant parameter is $m_{top}$, which also has a strong impact on $m_{H_u}^2$. In our runs we have used the value $m_{top} = 171.1$ GeV. However, we point out that the effect of taking a different value for $m_{top}$ can be compensated by choosing proper values for the angles $\theta_{1,2}$.

\subsection{Flavoured co-annihilation}

The $\Omega$LR model is in principle well suited for flavoured
co-annihilation \cite{Choudhury:2011um}. The $\tilde{\tau}_R$ can be
made lighter by flavour contributions, making it possible to have
co-annihilation with the lightest neutralino in points of parameter
space where it would be impossible without flavour effects.  Moreover,
to obtain the correct relic density one must include flavour violating
processes, like $\tilde{\tau}_R \tilde{\chi}_1^0 \to \gamma \mu$ or
$\tilde{\tau}_R \tilde{\tau}_R \to \tau \mu$.

\begin{figure}
\begin{center}
\vspace{5mm}
\includegraphics[width=0.47\textwidth]{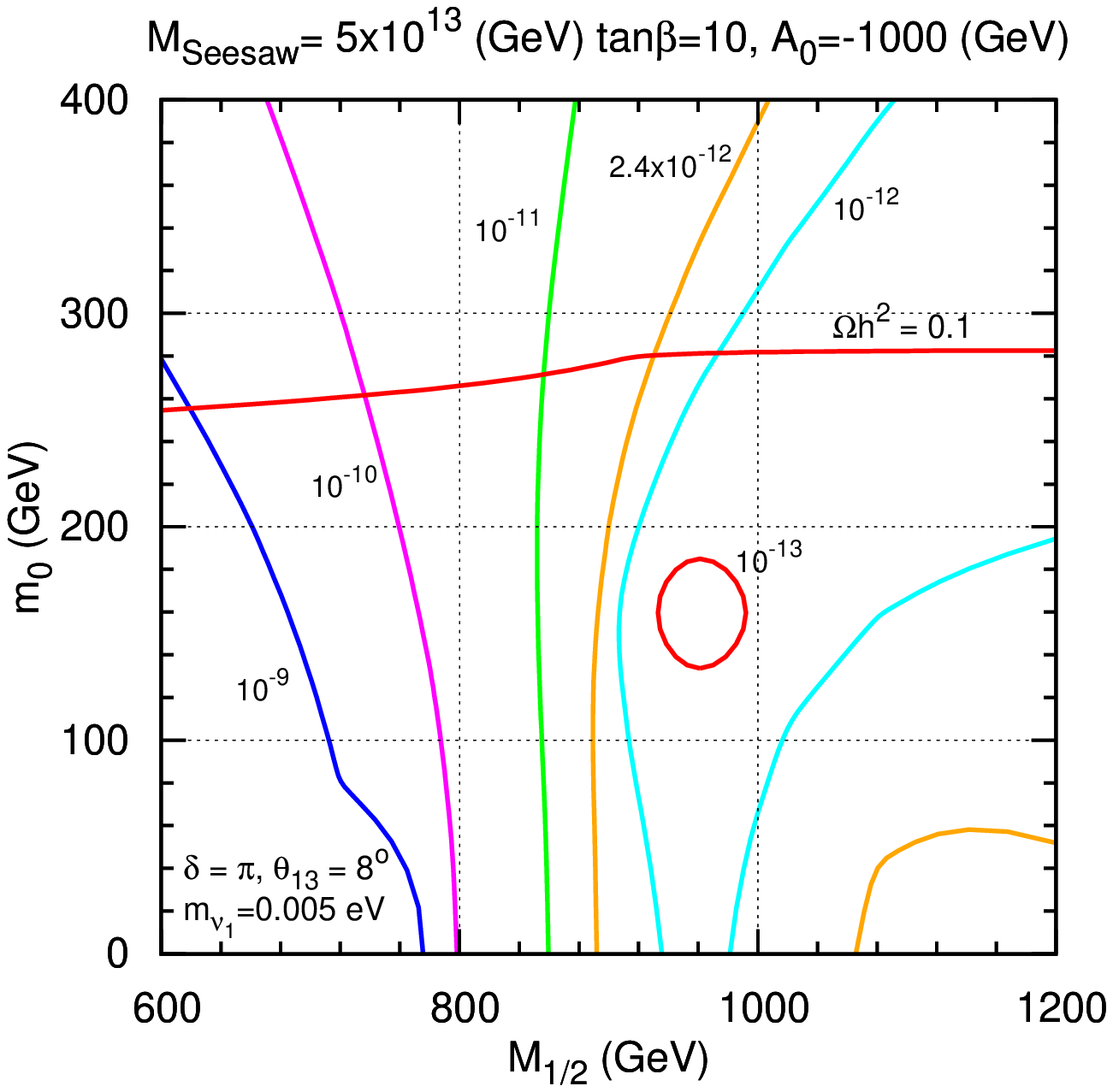}
\includegraphics[width=0.47\textwidth]{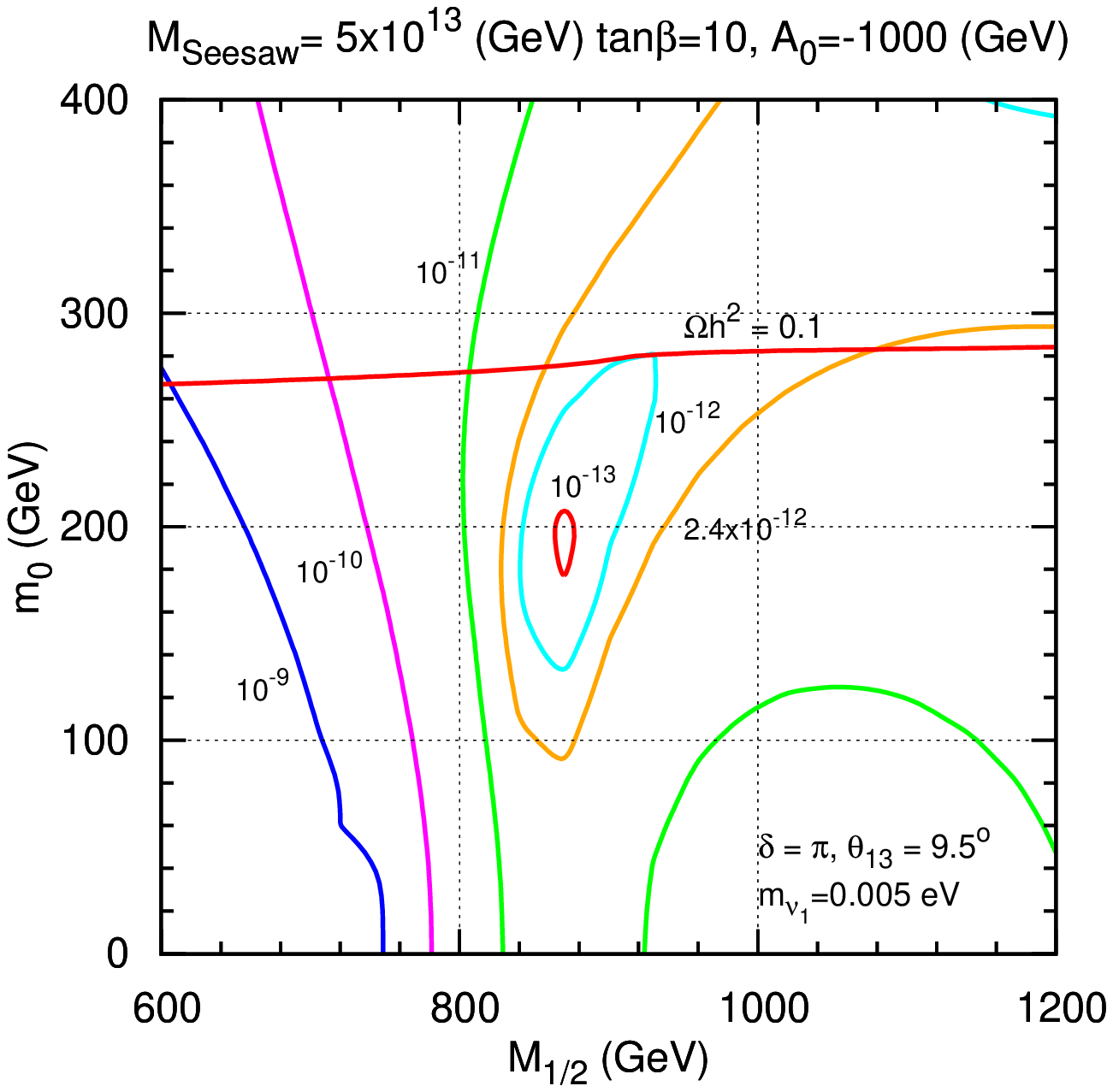}
\end{center}
\vspace{-5mm}
\caption{$\Omega_{\tilde\chi^0_1}h^2$ and $\text{Br}(\mu \to e \gamma)$ contour plots in the $m_0$-$M_{1/2}$ plane for $\delta=\pi$, $\theta_{13}=8^{0}$ (left) 
and $\theta_{13}=9.5^{0}$ (right). The other parameters are indicated
in the figures, except for the $\Omega$LR parameters, $v_R=10^{15}$ GeV
and $v_{BL}=10^{14} $ GeV.}
\label{fig:flav-coa-1}
\end{figure}

It is possible to find regions in the $m_0$-$M_{1/2}$ plane where
$\text{Br}(\mu \to e \gamma)$ and the other LFV decays respect the
experimental limits while having flavoured co-annihilation. In addition
to the usual fine-tuning that is required in the CMSSM to obtain the
observed dark matter relic density, flavoured co-annihilation also
requires to tune the neutrino mixing parameters 
in order to suppress $\text{Br}(\mu \to e
\gamma)$.  In order to have large flavour effects in the slepton
sector, to be able to reduce the mass of the $\tilde{\tau}_R$ and, at
the same time, respect the experimental limits, one must find values
for $\theta_{13}$ (the reactor angle) and $\delta$ (the Dirac phase) 
that allow cancellations in $\text{Br}(\mu \to e \gamma)$. 
We find that this cancellation is more effective if the mass of the
lightest neutrino is non-zero. 
The $\tau$ LFV decays are put under control by choosing
$m_0$ and $M_{1/2}$ sufficiently large.
Figure~\ref{fig:flav-coa-1} shows one example. On the left panel the
cancellation of $\text{Br}(\mu \to e \gamma)$ is obtained with $\delta
= \pi$ and $\theta_{13} = 8^{\circ}$, while on the right panel with
$\theta_{13} = 9.5^{\circ}$. Note that this occurs for different values of
$m_0$ and $M_{1/2}$.

Let us comment on some particularities of the $\Omega$LR model
concerning cancellations in $\text{Br}(\mu \to e \gamma)$. It turns
out that, contrary to the minimal type-I seesaw model, where the
cancellation in $\text{Br}(\mu \to e \gamma)$ occurs for a Dirac phase
$\delta = \pi$ and for $\theta_{13} \simeq 5^{\circ}$, independently
of the $m_0$ and $M_{1/2}$ values\footnote{See reference
\cite{Hirsch:2008dy} for examples.}, for the $\Omega$LR model the
cancellation occurs for different values of $\theta_{13}$ depending on
$m_0$ and $M_{1/2}$. This fact might be puzzling, since the Yukawa
structure that appears in the running of $m_L^2$ and $m_e^2$ is the
same as in the type-I seesaw. Therefore, if one wants to cancel this
combination of Yukawas by choosing the right $\theta_{13}$, it should
be possible to use the same value in type-I and $\Omega$LR, and this
is not the case.

The key point is that, in addition to $m_L^2$ and $m_e^2$, one also
has a soft trilinear term contribution, $T_e$, to $\text{Br}(\mu \to e
\gamma)$. This contribution is present in both amplitude coefficients,
$A_L$ and $A_R$, and suppressed in the type-I seesaw due to its
proportionality to charged lepton masses. However, in the $\Omega$LR
model this contributions turns out to be much larger. This can be
understood from the RGEs.  The 1-loop RGEs for $T_e$ in the parity
conserving regime \cite{Staub:Web} contain terms of the type
$f^\dagger f Y_L^{(k)}$ with $k = 1,2$. This will lead to two pieces
after applying the matching conditions at $v_R$, see
eq.~\eqref{matching-2}: the conventional $f^\dagger f Y_e$, which is
charged lepton mass suppressed, and the new $f^\dagger f Y_\nu$, which
is not suppressed and gives rise to new contributions to
$\text{Br}(\mu \to e \gamma)$.

This conclusion has been checked numerically, by studying the
dependence of the different contributions to $\text{Br}(\mu \to e
\gamma)$ on $\theta_{13}$. We found that for large $A_0$ the $T_e$ contribution
is dominant, which explains why $\text{Br}(\mu \to e \gamma)$ does not
cancel for $\theta_{13} \simeq 5^{\circ}$ as in the minimal type-I seesaw. For $A_0 = 0$ the $T_e$
contribution is much smaller, and thus one finds an approximate
cancellation for $\theta_{13} \simeq 5^{\circ}$, as expected.\\

\section{LHC phenomenology}
\label{sec:lhc}
In this section we investigate predictions for the LHC phenomenology of
the $\Omega$LR model.
The effects of the heavy states are two-fold:
they change the spectrum and induce flavour violating off-diagonal
elements in the slepton and sneutrino mass matrices. The changes
of the spectrum can
 lead to a potentially measurable difference
between the states which are mainly selectron- and smuon-like 
\cite{Allanach:2008ib,Buras:2009sg,Abada:2010kj,Esteves:2010si}. 
Here we are going to focus on the lepton flavour violation
within the SUSY cascade decays as they occur for example
in the decay chain
\begin{eqnarray}
\tilde q_L \to q \tilde \chi^0_2 \to q e^- {\tilde l}_i^+
\to q  e^- \mu^+ \tilde \chi^0_1
\end{eqnarray}
As already stated above, 
in this model one has additional lepton flavour mixing
for the R-sleptons as opposed to the usual seesaw
mechanisms where the lepton flavour mixing is in the left sector
only.

\begin{figure}[t]
\includegraphics[width=0.49\textwidth]{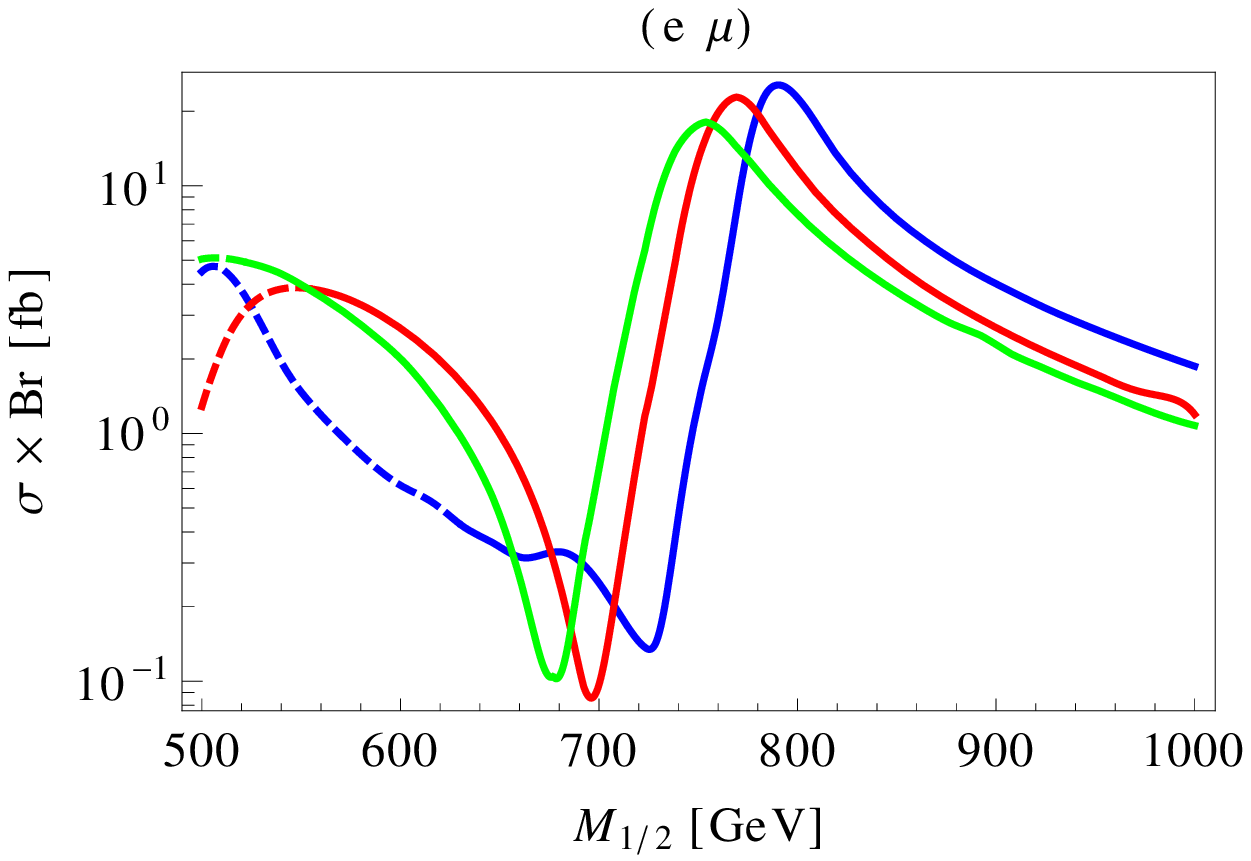}
\includegraphics[width=0.49\textwidth]{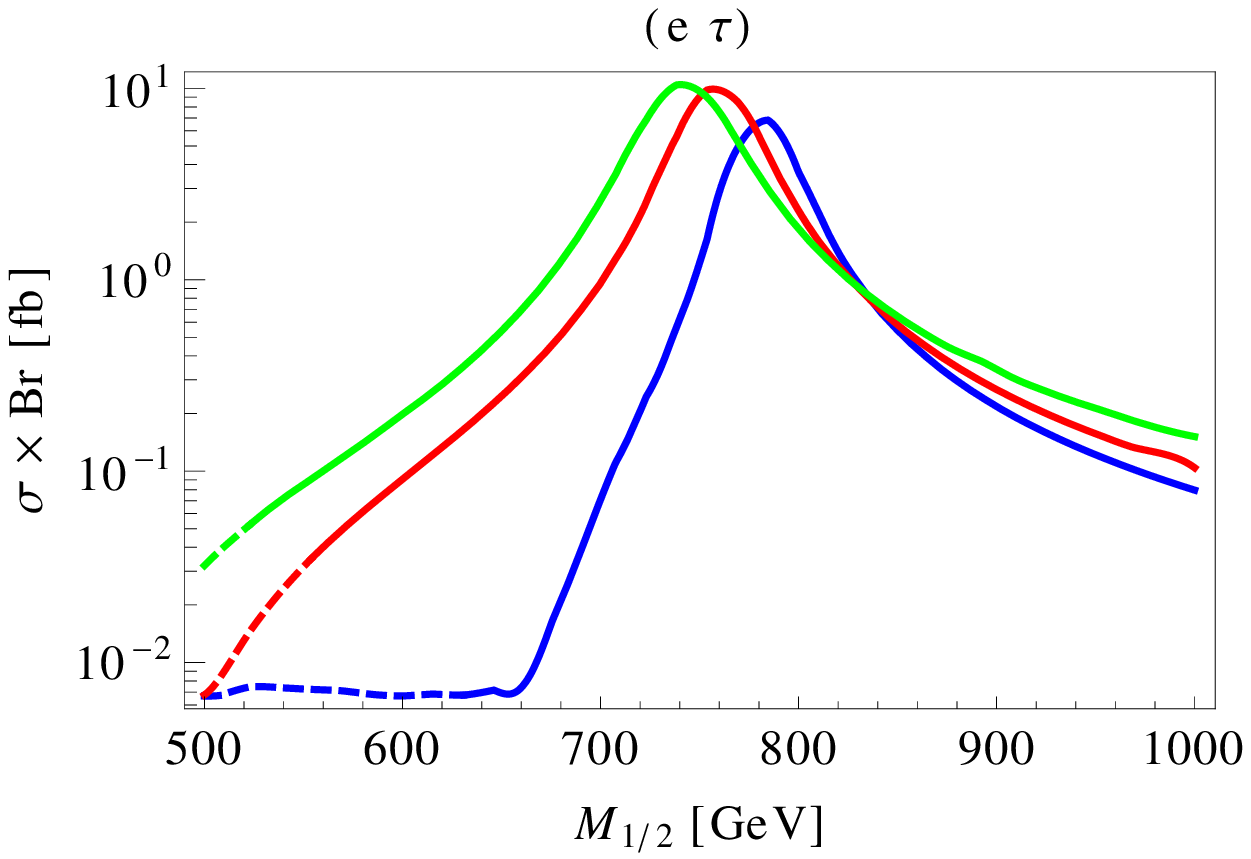}  \\[2mm]
\includegraphics[width=0.49\textwidth]{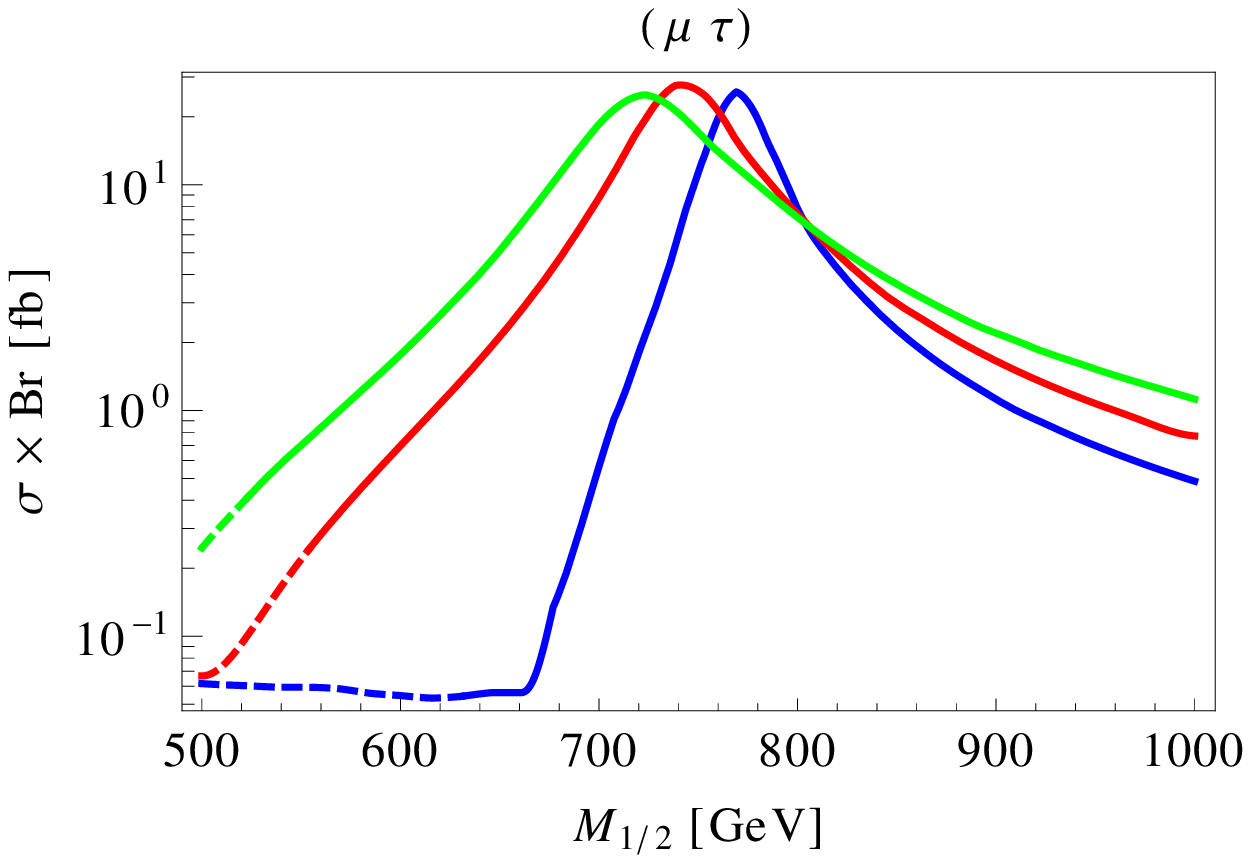} 
\caption{$\sigma \times \text{Br}$ for 
$pp \to \tilde{\chi}_2^0 \: X \to \tilde{\chi}_1^0 \: e \: \mu \: X$ 
(upper left plot), 
$pp \to \tilde{\chi}_2^0 \: X \to \tilde{\chi}_1^0 \: e \: \tau \: X$ 
(upper right plot), 
$pp \to \tilde{\chi}_2^0 \: X \to \tilde{\chi}_1^0 \: \mu \: \tau \: X$ 
(lower plot), 
as a function of $M_{1/2}$ for $\sqrt{s}=14$~TeV. 
The curves correspond to different values for $v_{BL} = v_R$: 
$10^{14}$ GeV (blue), $5 \times 10^{14}$ GeV (red) and $10^{15}$ GeV 
(green). The other CMSSM parameters have been fixed to: $m_0 = 100$ GeV,
 $A_0 = 0$ GeV, $\tan \beta = 10$ and $\mu > 0$. The seesaw scale $M_S$ has 
 been adjusted at each point in order to get 
 $\text{Br}(\mu \to e \gamma)$ close to but smaller than $2.4 \cdot 10^{-12}$. 
 Regions with a gluino mass below $1$ TeV are denoted with dashed lines.}
\label{fig:sigmabrm12emu}
\end{figure}

In Figure~\ref{fig:sigmabrm12emu}  we
show total rates for $pp \to \tilde \chi^0_2 \to l l' \tilde \chi^0_1$
with $l \ne l'$ as a function of $M_{1/2}$ for 
$\sqrt{s}=14$~TeV, $m_0=100$~GeV,
$A_0=0$, $\tan\beta=10$ and $\mu>0$ for different values
of $v_{BL}=v_R$. Choosing  $v_{BL}=v_R$ implies that one
has approximately the same amount of flavour mixing for left- and
right-sleptons.
Here we have summed over all initial states
containing squarks and gluinos and over all cascade decays
leading to a $\tilde \chi^0_2$. We have furthermore required that
exactly one $\tilde \chi^0_2$ decays lepton flavour violating and
that no additional lepton occurs in any of the cascade decays of
the corresponding event. For the calculation of the cross section
we have used the {\tt FASER-LHC} package \cite{BenFaser}
which is based on the program {\tt PROSPINO} \cite{Beenakker:1996ed}. 
Recent ATLAS \cite{daCosta:2011qk} and CMS \cite{Chatrchyan:2011zy}
data constrain already gluino and squark masses and are mainly
interpreted in the context of the CMSSM. In our model the spectrum 
differs with respect to the CMSSM but we expect that for our choice
of parameters gluino masses are excluded below about 1 TeV. For this reason
we put dashed lines if the gluino mass is below this bound and full
lines if the value is above.

The neutrino Yukawa couplings are chosen such that Br$(\mu\to e \gamma)$
is close to its current bound of $2.4 \cdot 10^{-12}$. 
Here we assumed that the right-handed neutrinos are
degenerate in mass.
Let us first discuss the final states containing a $\tau$-lepton.
We see an increase of the rates with increasing $M_{1/2}$ until
it reaches a maximum between 700 and 800 GeV depending on the value
$v_{BL}=v_R$ and then drops with increasing $M_{1/2}$. This behaviour
is due to an interplay of three effects: (i) with increasing $M_{1/2}$
larger values for the entries of $Y_\nu$ are allowed as 
Br($\mu\to e \gamma$) gets suppressed by the heavier spectrum.
This implies larger flavour violating decay rates of $\tilde \chi^0_2$.
(ii) For fixed $m_0$ the left sleptons are heavier than 
$\tilde \chi^0_2$ for small $M_{1/2}$ and they have about the same
mass for $M_{1/2}$ in our examples between 500 and 600 GeV depending
on the value of   $v_{BL}=v_R$.
(iii) Heavier squark and gluino masses imply reduced cross sections.
The shift of the maxima to higher values of $M_{1/2}$ with decreasing
$v_{BL}=v_R$ is again a consequence of the modified spectrum:
smaller values of $v_{BL}=v_R$ imply smaller gaugino masses
and increased ratios of slepton masses over gaugino masses at the
electroweak scale for fixed $M_{1/2}$, which shifts the $M_{1/2}$ value
where neutralino decays into on-shell L-sleptons are kinematically
allowed. For completeness we note that
the total rates can go up to 10 (30) fb in case of the $e \tau$
($\mu\tau$) implying at most a few thousands 
events if the design luminosity
of 100 fb$^{-1}$ per year can be achieved. The reduced values for the
$e \tau$ channel compared to the $\mu\tau$ is a consequence
of neutrino
physics as we require tri-bimaximal mixing. This leads to
$m^2_{L,13}$/$m^2_{L,23} \simeq m^2_{e^c,13}$/$m^2_{e^c,23} \sim
m_{sol}/m_{atm}$ where $m_{sol}$ and $m_{atm}$ are the solar and
atmospheric neutrino mass scale, respectively.

In case of the $e$-$\mu$ final state in Figure~\ref{fig:sigmabrm12emu}
we see a second maximum at lower values of $M_{1/2}$ in a region
where the final states containing a $\tau$ have rates which
are two to three orders of magnitude smaller. The reason for this
is a level crossing of the $\tilde e_L$ with $\tilde \mu_L$ leading
to an enhancement of three body decays 
$\tilde \chi^0_2 \to e \mu \tilde \chi^0_1$ via virtual left sleptons.
At the maxima we find that
Br($\tilde \chi^0_2 \to e^+ e^- \tilde \chi^0_1) \simeq $
Br($\tilde \chi^0_2 \to e^+ \mu^- \tilde \chi^0_1) \simeq $
Br($\tilde \chi^0_2 \to \mu^+ \mu^- \tilde \chi^0_1) \simeq $
Br($\tilde \chi^0_2 \to \tau^+ \tau^- \tilde \chi^0_1)/2$.
These three body decays give the dominant contribution to the flavour
violating signal.

Finally let us comment on the impact of the parameters we have kept
fixed so far:
for moderate increases, $m_0$ shifts the maximum 
of the rates as the mass
of the sleptons is increased. For sufficiently large $m_0$ two
body decays into sleptons become kinematically forbidden implying
negligible event rates. A variation of $A_0$ has only a small
impact as it leads mainly to a shift of the masses for the states
which are mainly stau-like.  Larger values of $\tan\beta$ reduce
the lepton flavour violating signal as Br$(\mu\to e \gamma)$ grows
like $\tan^2\beta$ implying smaller allowed values for the flavour
off-diagonal entries in the slepton mass matrices and, thus, reduced
branching ratios for the lepton flavour violating neutralino decays. 

\begin{figure}[t]
\begin{center}
\vspace{5mm}
\includegraphics[width=0.5\textwidth]{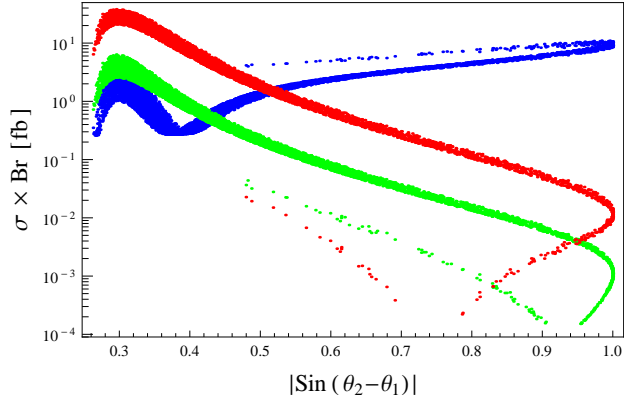}
\end{center}
\vspace{-5mm}
\caption{$\sigma \times \text{Br}$ for $pp \to \tilde{\chi}_2^0 \: X
\to \tilde{\chi}_1^0 \: l \: l' \: X$ as a function of 
$|\sin(\theta_2 - \theta_1)|$. The bands correspond to the following
$l,l'$ combinations: $e$ -$\mu$ (blue), $e$-$\tau$ (green) and 
$\mu$-$\tau$ (red).
The parameters have been taken as in Figure~\ref{fig:randtheta}. A center of mass energy of $14$ TeV has been assumed.}
\label{fig:sigmabrtheta21A}
\end{figure}

\begin{figure}[t]
\begin{center}
\vspace{5mm}
\includegraphics[width=0.5\textwidth]{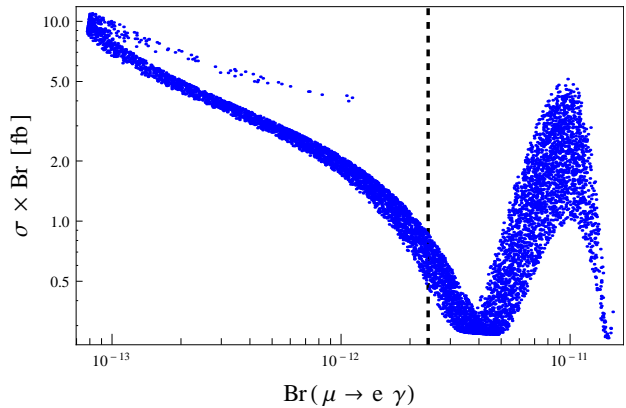}
\end{center}
\vspace{-5mm}
\caption{$\sigma \times \text{Br}$ for $pp\to \tilde{\chi}_2^0 \:X \to
\tilde{\chi}_1^0 \: e \: \mu \: X$ for $\sqrt{s}=14$~TeV as a function
of Br($\mu \to e \gamma$) The parameters have been taken as in
Figure~\ref{fig:randtheta}. The vertical line shows the current MEG bound
$\text{Br}(\mu \to e \gamma) < 2.4 \cdot 10^{-12}$.}
\label{fig:sigmabrtheta21B}
\end{figure}

Up to now we have fixed the parameters such that Br$(\mu \to e \gamma)$
is close to its experimental bound. In the following we depart from
this by performing a scan varying in particular 
$|\sin(\theta_2 - \theta_1)|$ because a strong dependence of the 
rare lepton decays on
this quantity has been found, see Figure~\ref{fig:randtheta}. 
In Figure~\ref{fig:sigmabrtheta21A} we show the rates
for $pp \to \tilde{\chi}_2^0 \: X
\to \tilde{\chi}_1^0 \: l \: l' \: X$ as a function of 
$|\sin(\theta_2 - \theta_1)|$ for $\sqrt{s}=14$~TeV and fixing
the parameters as in Figure~\ref{fig:randtheta}. The final states
containing a $\tau$-lepton behave similar as the corresponding
rare $\tau$ decays. However, in case of the $e$-$\mu$ final state
we find a lower limit of about 0.1 fb for this parameter set. 
The reason is, that
above $|\sin(\theta_2 - \theta_1)|\simeq 0.3$ the mass
splitting between the selectrons and smuons get smaller the larger 
$|\sin(\theta_2 - \theta_1)|$ is. This holds for both, left and right
states. This leads to a decrease of Br$(\mu \to e \gamma)$ and at
the same time to a slight increase of the lepton flavour violating signals
at the LHC as can also be seen in Figure~\ref{fig:sigmabrtheta21B}
where we show the LHC rate of the $e\mu$ final state as a function
of Br$(\mu \to e \gamma)$.

In principle one might worry about the fact that the mass differences
between the sleptons are partly smaller than the corresponding widths 
implying potential resonance effects 
\cite{ArkaniHamed:1996au,ArkaniHamed:1997km}. We have checked that the
principal features discussed above
remain if one does not use the narrow width approximation,
leading to  the cascades 
$\tilde \chi^0_2 \to l \tilde l_j \to l l'\tilde \chi^0_1$,
but calculates the complete amplitudes using Breit-Wigner propagators
for the sleptons. This shifts the
results only very slightly. 

\section{Conclusions}

We have studied phenomenological aspects of a supersymmetric 
left-right model. Our discussion has centered on two aspects: 
(a) Changes in the low energy SUSY spectra with respect to 
CMSSM expectations and the resulting consequences for the 
relic density of the lightest neutralino, assumed to be the 
cold dark matter of the universe. And (b) Lepton flavour 
violation induced by the LFV entries in the neutrino Yukawa 
coupling matrix, required to explain the observed neutrino 
angles, and consequences for the LHC.

In the CMSSM the lightest neutralino can have a relic density 
in agreement with the measured value \cite{Nakamura:2010zzi} 
only in some very specific parts of parameter space. These 
well-known regions are (i) the stau co-annihilation region; 
(ii) stop co-annihilation region; (iii) the focus point line 
and (iv) the Higgs funnel. The modified running of the 
soft parameters in the $\Omega$LR model shifts the allowed regions. 
In general, the lower $v_{BL}$ and 
$v_R$ are, the smaller these regions become until for certain 
values of these parameters (depending on the specific region) 
the DM allowed regions disappear completely. From an observation 
of a SUSY spectrum consistent with any of these regions one could 
infer lower limits on $v_{BL}$ and $v_R$ not far below the GUT scale 
for CMSSM boundary conditions.

The model also allows for flavoured co-annihilation 
\cite{Choudhury:2011um}. We have given some specific examples of 
parameter points in which it is possible to have large flavoured 
contributions to the co-annihilation cross section, despite the 
stringent upper limit on the decay $\mu\to e \gamma$ recently 
published by the MEG collaboration \cite{Adam:2011ch}. 

We have shown that LFV decays of the $\tilde\chi^0_2$ can reach 
up to 20 fb$^{-1}$ for the LHC at $\sqrt{s}=14$~TeV. All combinations 
of different lepton flavour final states could be large. These 
as well as the rare lepton decays show a strong dependence on 
the parameters of the model, noteworthy also on the high scale 
parameters $v_R$ and $\sin(\theta_2-\theta_1)$.

\section*{Acknowledgements}

We thank B.~O'Leary for providing his package {\tt FASER-LHC}.
W.P.~thanks the IFIC for hospitality during an extended stay
and the Alexander von Humboldt foundation for financial support.
F.S, W.P.\ and A.V.\ have been supported by the DFG, project
number PO-1337/1-1.
We acknowledge support from the Spanish MICINN grants FPA2008-00319/FPA, 
FPA2008-04002/E and MULTIDARK CAD2009-00064 and
by the Generalitat Valenciana grant Prometeo/2009/091 and the 
EU~Network grant UNILHC PITN-GA-2009-237920. J.N.E and J.C.R~also acknowledge
support from  {\it Funda\c{c}\~ao para a Ci\^encia e a Tecnologia},
grants CFTP-FCT Unit 777 and CERN/FP/116328/2010. Moreover, J.N.E. and A.V. were also
partially suported by Marie Curie Early Initial Training Network
Fellowships of the European Community's Seventh Framework Programme under contract
number (PITN-GA-2008-237920-UNILHC).

\appendix

\section{Invariants} \label{ap:inv}

Neglecting the contributions from the Yukawa couplings and working in the leading-log approximation, the soft mass parameters at the SUSY scale are computed to be
\begin{equation}
M_i(m_{SUSY}) = \frac{\alpha_i(m_{SUSY})}{\alpha_{BL}(m_{GUT})} M_{1/2}
\end{equation}
\begin{equation} \label{mfsoft}
m_f^2 = m_0^2 + \frac{M_{1/2}^2}{\alpha_{BL}^2(m_{GUT})} \left[ \alpha_i^2(v_{BL}) f_{MSSM}^i + \check{\alpha}_i^2(v_R) f_{BL}^i + \alpha_i^2(m_{GUT}) f_{LR}^i \right]
\end{equation}
In equ.~\eqref{mfsoft} a sum over the index $i$ is implied, with $i=(Y,2,3)$ in the first and second terms and $i=(BL,2,3)$ in the third\footnote{In order to avoid any possible confusion at $\mu = v_R$, we use the notation $\check{\alpha}_i(v_R) \equiv (\alpha_Y,\alpha_2,\alpha_3)(v_R)$ for the couplings right below $v_R$ and $\hat{\alpha}_i(v_R) \equiv (\alpha_{BL},\alpha_2,\alpha_3)(v_R)$ for the couplings right above $v_R$.}. The $f^i$ parameters are defined as
\begin{eqnarray}
f_{MSSM}^i &=& \frac{B_i^{\tilde{f}}}{b_{MSSM}^i} \left[ 1 - \left(\frac{\alpha_i(m_{SUSY})}{\alpha_i(v_{BL})}\right)^2 \right] \label{fMSSM} \\
f_{BL}^i &=& \frac{B_i^{\tilde{f}}}{b_{BL}^i} \left[ 1 - \left(\frac{\alpha_i(v_{BL})}{\check{\alpha}_i(v_R)}\right)^2 \right] \label{fBL} \\
f_{LR}^i &=& \frac{A_i^{\tilde{f}}}{b_{LR}^i} \left[ 1 - \left(\frac{\hat{\alpha}_i(v_R)}{\alpha_i(m_{GUT})}\right)^2 \right] \label{fLR}
\end{eqnarray}

For the numerical studies we have fixed the  GUT scale to 
be $m_{GUT} = 2 \cdot 10^{16}$ GeV. The gauge couplings in the previous equations are determined by means of the general formula
\begin{equation}
\alpha_i(\mu_2) = \alpha_i(\mu_1) \left[ 1 - \frac{\alpha_i(\mu_1)}{4 \pi} b^i \ln \left( \frac{\mu_2^2}{\mu_1^2} \right) \right]^{-1}
\end{equation}
using the experimental values at $\mu = m_Z$ as starting point. In addition, at $\mu = v_R$ one must apply the following matching conditions
\begin{eqnarray}
\hat{\alpha}_{BL}^{-1} &=& \frac{5}{2} \check{\alpha}_Y^{-1} - \frac{3}{2} \check{\alpha}_2^{-1} \\
\hat{\alpha}_2 &=& \check{\alpha}_2 \\
\hat{\alpha}_3 &=& \check{\alpha}_3
\end{eqnarray}
In the previous three equations, the gauge couplings on the right-hand side are the ones for the $\mu < v_R$ regime, whereas the gauge couplings on the left-hand side are the ones for the $\mu > v_R$ regime.

The beta coefficients used in the previous formulas are
\begin{eqnarray}
b^{SM} &=& (b_Y^{SM},b_2^{SM},b_3^{SM}) = (\frac{41}{10},-\frac{19}{6},-7) \\
b^{MSSM} &=& (b_Y^{MSSM},b_2^{MSSM},b_3^{MSSM}) = (\frac{33}{5},1,-3) \\
b^{BL} &=& (b_Y^{BL},b_2^{BL},b_3^{BL}) = (\frac{33}{5},3,-3) \\
b^{LR} &=& (b_1^{LR},b_2^{LR},b_3^{LR}) = (24,8,-3)
\end{eqnarray}
Finally, the $A_i^{\tilde{f}}$ and $B_i^{\tilde{f}}$ coefficients are
\renewcommand{\tabcolsep}{1cm}
\renewcommand{\arraystretch}{2.2}
\begin{center}
\begin{tabular}{c|c c c c c}
 & $\displaystyle \tilde{e}^c$ & $\displaystyle \tilde{L}$ & $\displaystyle \tilde{d}^c$ & $\displaystyle \tilde{u}^c$ & $\displaystyle \tilde{Q}$ \\
\hline
$A_{BL}$ & $\displaystyle \frac{3}{4}$ & $\displaystyle \frac{3}{4}$ & $\displaystyle \frac{1}{12}$ & $\displaystyle \frac{1}{12}$ & $\displaystyle \frac{1}{12}$ \\
$A_2$ & $\displaystyle \frac{3}{2}$ & $\displaystyle \frac{3}{2}$ & $\displaystyle \frac{3}{2}$ & $\displaystyle \frac{3}{2}$ & $\displaystyle \frac{3}{2}$ \\
$A_3$ & $0$ & $0$ & $\displaystyle \frac{8}{3}$ & $\displaystyle \frac{8}{3}$ & $\displaystyle \frac{8}{3}$ \\
$B_{Y}$ & $\displaystyle \frac{6}{5}$ & $\displaystyle \frac{3}{10}$ & $\displaystyle \frac{2}{15}$ & $\displaystyle \frac{8}{15}$ & $\displaystyle \frac{1}{30}$ \\
$B_{2}$ & $0$ & $\displaystyle \frac{3}{2}$ & $0$ & $0$ & $\displaystyle \frac{3}{2}$ \\
$B_{3}$ & $0$ & $0$ & $\displaystyle \frac{8}{3}$ & $\displaystyle \frac{8}{3}$ & $\displaystyle \frac{8}{3}$
\end{tabular}
\end{center}

\end{document}